\documentclass[twocolumn,showpacs,nofootinbib,preprintnumbers,amsmath,amssymb,showkeys,superscriptaddress]{revtex4}
\usepackage{epsfig}
\usepackage{amssymb}
\usepackage{amsmath}
\usepackage{graphicx}
\usepackage{bm}
\usepackage{color}

\newcommand{\beq}{\begin{equation}}
\newcommand{\eeq}{\end{equation}}
\newcommand{\bea}{\begin{eqnarray}}
\newcommand{\eea}{\end{eqnarray}}
\newcommand{\nn}{\nonumber \\}

\newcommand\eqn[1]{(\ref{#1})}      
\newcommand\Eqn[1]{Eq.~(\ref{#1})}  

\newcommand{\tr}{\hbox{tr}}
\newcommand{\bartheta}{\,\bar{\!\theta}}
\newcommand{\thetab}{{\bartheta}}
\newcommand{\tb}{{\bartheta}}
\newcommand{\ts}{{\theta}}
\newcommand{\cb}{{\bar c}}

\begin{document}

\title{Covariant gauges without Gribov ambiguities in Yang-Mills theories}

\author{J. Serreau}
\affiliation{Astro-Particule et Cosmologie (APC), CNRS UMR 7164, Universit\'e Paris 7 - Denis Diderot\\ 10, rue Alice Domon et L\'eonie Duquet, 75205 Paris Cedex 13, France.}
\author{M. Tissier}
\affiliation{LPTMC, Laboratoire de Physique Th\'eorique de la Mati\`ere Condens\'ee, CNRS UMR 7600, Universit\'e Pierre et Marie Curie, \\ boite 121, 4 place Jussieu, 75252 Paris Cedex 05, France}
\author{A. Tresmontant}
\affiliation{Astro-Particule et Cosmologie (APC), CNRS UMR 7164, Universit\'e Paris 7 - Denis Diderot\\ 10, rue Alice Domon et L\'eonie Duquet, 75205 Paris Cedex 13, France.}
\affiliation{LPTMC, Laboratoire de Physique Th\'eorique de la Mati\`ere Condens\'ee, CNRS UMR 7600, Universit\'e Pierre et Marie Curie, \\ boite 121, 4 place Jussieu, 75252 Paris Cedex 05, France}

\date{\today}

\begin{abstract}
We propose {  a one-parameter family of nonlinear covariant gauges which can be formulated as an extremization procedure that may be amenable to lattice implementation}. At high energies, where the Gribov ambiguities can be ignored, this reduces to the Curci-Ferrari-Delbourgo-Jarvis gauges. We further propose a continuum formulation in terms of a local action which is free of Gribov ambiguities and avoids the Neuberger zero problem of the standard Faddeev-Popov construction. This involves an averaging over Gribov copies with a nonuniform weight, which introduces a new gauge-fixing parameter. We show that the proposed gauge-fixed action is perturbatively renormalizable in four dimensions and we provide explicit expressions of the renormalization factors at one loop. We discuss the possible implications of the present proposal for the calculation of Yang-Mills correlators.
\end{abstract}

\pacs{11.15.-q, 12.38.Bx}
\keywords{Yang-Mills theories, gauge-fixing, Gribov ambiguities, infrared correlation functions}

\maketitle

\section{Introduction}\label{sec:introduction}

The understanding of the long distance properties of non-Abelian gauge theories is a 
problem of topical importance. Perturbation theory breaks down at low energies since the running coupling constant increases without bound. The highly nontrivial infrared dynamics of Yang-Mills (YM) fields is thus thought to be accessible only 
through nonperturbative techniques. Among existing such approaches, only lattice 
calculations can directly access physical observables. In contrast, continuum methods, such as truncations of Schwinger-Dyson equations \cite{vonSmekal97,Alkofer00} or  the nonperturbative renormalization group \cite{Ellwanger96}, are based on computing the basic correlation functions of Yang-Mills fields and require a gauge-fixing procedure. It is thus of key importance to have a quantitative understanding of such correlators. 

When possible, gauge-fixed lattice calculations provide an important benchmark for continuum approaches. An important issue concerns the algorithmic complexity of fixing a gauge numerically. For instance, covariant gauges, which are the most convenient for continuum calculations, are not easily implemented on the lattice as they require one to find the roots of a large set of coupled nonlinear equations. The Landau gauge is a remarkable exception because it can be formulated as a minimization procedure well suited to numerical methods and which, for this reason, has been extensively studied \cite{Cucchieri_08b,Cucchieri_08,Bornyakov2008,Bogolubsky09,Dudal10,Maas:2011se}. Precise determinations of the ghost-antighost and gluon two-point correlators in this gauge have now been obtained. In particular, these show the so-called decoupling behavior in the infrared, where both the gluon correlator and the ghost dressing function are finite at zero mo\-men\-tum~\cite{Boucaud06,Aguilar07,Aguilar08,Boucaud08,Fischer08,RodriguezQuintero10,Boucaud:2011ug,Huber:2012kd}. 

However, the Landau gauge is a peculiar representative of the class of covariant gauges as it possesses additional symmetries. It is of great interest to investigate other gauges within both lattice and continuum approaches in order to distinguish the specific features of the Landau gauge from more generic ones as well as to study the gauge dependence---and thus the possible gauge independent features---of Yang-Mills correlators.\footnote{For analytical studies of the infrared gluon and ghost two-point correlators including nonperturbative features, see, e.g., \cite{Dudal:2003by,Sobreiro:2005vn,Aguilar:2007nf}.} 
Attempts to formulate general linear covariant gauges on the lattice were made in \cite{Giusti:1996kf,Giusti:1999im} and, later, in \cite{Cucchieri:2008zx,Cucchieri:2009kk}, but these were not completely satisfactory. In particular, although the proposal of \cite{Cucchieri:2009kk} solves most of the problems afflicting the methods proposed earlier, it is limited to infinitesimal gauge transformations as a result of trying to enforce a linear gauge-fixing condition.

The proposals mentioned above rely on a suitable extremization procedure. An alternative strategy has been proposed in \cite{Parrinello:1990pm,Zwanziger:1990tn}, which is based on sampling each gauge orbit with a nontrivial measure. This was shown to be tractable in lattice simulations, although numerically demanding, and exploratory physical studies were performed  \cite{Fachin:1991pu,Henty:1996kv}. However, the corresponding continuum action appears to be nonrenormalizable \cite{Zwanziger:1990tn,Fachin:1993qg}, which makes a proper continuum limit of lattice calculations problematic \cite{Henty:1996kv}. To our knowledge, this has not been pursued further. 

The second, related issue concerning gauge fixing in non-Abelian theories is the existence of Gribov ambiguities \cite{Gribov77} for the most common choices of gauge, including covariant gauges. Continuum approaches are essentially based on the standard Faddeev-Popov procedure, which neglects Gribov copies but which is assumed to be a valid starting point at sufficiently high energies. On the lattice, the Faddeev-Popov construction is, however, plagued by the Neuberger zero problem \cite{Neuberger:1986vv}, due to the degenerate contribution of many Gribov copies with alternating signs. The easy way to cope with this issue is to pick up a single copy per gauge orbit. This is the essence of the so-called minimal Landau gauge \cite{Boucaud:2011ug}. However, such a procedure is not easy to formulate with continuum approaches, which complicates the task of comparing results. The Gribov-Zwanziger proposal \cite{Gribov77,Zwanziger89,Dell'Antonio:1991xt} to restrict the path integral to the first Gribov region is not sufficient since the latter is not free of Gribov ambiguities \cite{Vandersickel:2012tz}.\footnote{A refined version of the Gribov-Zwanziger scenario leads to predictions for the ghost and gluon two-point correlators which describe well the lattice data in the Landau gauge \cite{Dudal08}. An extension of the Gribov-Zwanziger scenario to linear gauges has been studied in \cite{Sobreiro:2005vn}.}

Recently, two of us proposed an alternative strategy in the case of the Landau gauge, namely, to average over Gribov copies in such a way as to lift their degeneracy in the Faddeev-Popov procedure and avoid the Neuberger zero problem \cite{Serreau:2012cg}.\footnote{A similar proposal has been made earlier in Ref. \cite{vonSmekal:2008en}. We thank L. von Smekal for pointing this to our attention. For other proposals addressing the Neuberger zero problem, see, e.g., Refs. \cite{Hughes:2012hg,vonSmekal:2013cla} and references therein.} This is somewhat similar to the proposal of \cite{Parrinello:1990pm,Zwanziger:1990tn} mentioned above, but where the average is restricted to Gribov copies along each gauge orbit and where we include a sign factor from the Faddev-Popov determinant.  Such an averaging procedure can be formulated in terms of a local action, suitable to continuum approaches and, for a proper choice of the weighting functional, the gauge-fixed theory is perturbatively renormalizable in four dimensions. This introduces a new gauge-fixing parameter, which controls the weight of the different copies. Remarkably, the resulting theory turns out to be perturbatively equivalent to a simple massive extension of the Landau gauge Faddeev-Popov action, namely the Curci-Ferrari (CF) model \cite{Curci:1976bt}, for what concerns the calculation of ghost and gluon correlators. This is an exciting result since a one-loop perturbative calculation in this model had been shown earlier to give a remarkably good description of lattice data down to the deep infrared regime \cite{Tissier_10}. In particular, the model was shown to possess infrared safe renormalization group trajectories, with no Landau pole.

The aim of the present paper is twofold and concerns the two issues mentioned above. In Sec.~\ref{sec_gaugefixing}, we express a one-parameter family of nonlinear covariant gauges as an extremization procedure {  valid for arbitrary, finite gauge transformations. The corresponding extremization functional generalizes the one of \cite{Cucchieri:2009kk} and presents good properties for the purpose of numerical extremization (minimization) techniques}. Neglecting Gribov ambiguities issues---which {  should be} justified at high energies---and implementing the standard Faddeev-Popov procedure leads to the Curci-Ferrari-Delbourgo-Jarvis (CFDJ) Lagrangian \cite{Curci:1976bt,Delbourgo:1981cm}. The latter is a perfectly valid gauge-fixed Lagrangian, with all good properties, including unitarity, but with Gribov ambiguities. 

In the second part of the paper, we extend the method of Ref. \cite{Serreau:2012cg} to deal with these Gribov copies. This involves a suitable averaging procedure along each gauge orbit, which we treat formally using the replica trick, borrowed from the theory of disordered systems in statistical physics \cite{young}. This allows us to formulate our gauge-fixing procedure in terms of a local action. The resulting gauge-fixed theory admits an elegant and powerful superfield description. It describes a set of replicated supersymmetric nonlinear sigma models coupled to a massive extension of the CFDJ Lagrangian, the general CF  Lagrangian \cite{Curci:1976bt}. In contrast to the case of the Landau gauge studied in \cite{Serreau:2012cg}, we find that the nonlinear sigma model sector does not decouple in that case, leading to explicit differences with the CF model. This is presented in Sec.~\ref{sec_averaging}. We analyze the symmetries of our gauge-fixed Lagrangian and prove its perturbative renormalizability in four dimensions to all orders in Sec.~\ref{sec_renorm}. We then derive the Feynman rules of the theory in Sec.~\ref{sec_frules} and we compute explicitly the renormalization factors at one-loop order in Sec.~\ref{sec_renor1loop}. 

We emphasize that the extremization {  functional proposed} in Sec.~\ref{sec_gaugefixing} is a slight generalization of the one routinely employed for the Landau gauge. It {  is thus interesting to investigate its numerical implementation by means of existing techniques, e.g.,  along the lines of Refs. \cite{Cucchieri:2009kk,Cucchieri:2010ku,Cucchieri:2011pp}. We hope the present paper will motivate such studies.} We expect that for some value of the weighting parameter, the average over Gribov copies described in Sec.~\ref{sec_averaging} is essentially equivalent to picking up a random copy, as in the minimal Landau gauge. In that case, our proposal predicts specific features for the basic Yang-Mills correlators. For instance, we expect the ghost correlator to develop a mass gap at vanishing momentum, as discussed in Sec.~\ref{sec_sum}, which may improve the infrared properties of perturbation theory. We also briefly mention in Sec.~\ref{sec_sum} specific two-point correlators that can be computed by numerical and analytical means and which may carry interesting information concerning the role of Gribov copies.

Some technical details and additional material are presented in the Appendices. { We discuss some aspects of the issue of numerical minimization in Appendix~\ref{appsec:min}. In particular, we show how the standard Los Alamos minimization algorithm can be straightforwardly generalized to the present proposal.} Appendix \ref{appsec:replica} shows how to exploit fully the replica symmetry relevant to our proof of renormalizability. In Appendix \ref{appsec_AT}, we describe an alternative formulation of our proposal which does not use the superfield formalism. We provide explicit one-loop results in this context and discuss in detail the relation to the superfield formalism, which involves composite field renormalization. Finally, we discuss the specific features of the Landau gauge in Appendix \ref{appsec_Landau}.

\section{The gauge-fixing procedure}
\label{sec_gaugefixing}

The classical action of the SU($N$) Yang-Mills theory reads, in $d$-dimensional Euclidean space,
\beq 
 S_{\rm YM}[A]=\frac{1}{4}\int_x \left(F_{\mu\nu}^a\right)^2\,, 
\eeq 
where $\int_x\equiv\int d^dx$ and 
\beq
 F_{\mu\nu}^a=\partial_\mu A_\nu^a-\partial_\nu A_\mu^a +g_0f^{abc}A_\mu^bA_\nu^c,
\eeq
where $g_0$ is the (bare) coupling constant and a summation over spatial and color indices is understood. In the following we use the convention that fields written without an explicit color index are contracted with the generators $t^a$ of SU($N$) in the fundamental representation and are thus $N\times N$ matrix fields, e.g. $A_\mu=A_\mu^a t^a$.  Our normalization for the generators is such that 
\begin{equation}
  \label{eq_prod_gene}
  t^a t^b=\frac{\delta ^{ab}}{2N}\openone +\frac{if^{abc}+d^{abc}}2 t^c,
\end{equation}
with $f^{abc}$ and $d^{abc}$, the usual totally antisymmetric and totally symmetric tensors of SU($N$). In particular, we have
\beq
  \tr \left(t^a t^b\right)=\frac{\delta^{ab}}{2}.
\eeq

In order to fix the gauge, we consider the functional
\begin{equation}
  \label{eq_func}
  {\cal H}[A,\eta,U]=\int_x\,\tr\left[\left(A^U_\mu\right)^2+ \frac{U^\dagger \eta+\eta^\dagger U}2\right]
\end{equation}
for each field configuration $A_\mu$, where $\eta$ is an arbitrary $N\times N$ matrix field and 
\begin{equation}
  \label{eq_gauge_transfo}
  A_\mu^U=UA_\mu U^\dagger+\frac i {g_0}U\partial_\mu U^\dagger
\end{equation}
is the gauge transform of $A_\mu$ with a gauge element $U\in$ SU($N$). We define our gauge condition as (one of) the extrema of ${\cal H}$ with respect to $U$. The latter can be obtained by writing $U\to V U$ with $V=e^{ig_0\lambda}$ and expanding in $\lambda$. Using $A_\mu^V= A_\mu+D_\mu\lambda+\mathcal O(\lambda^2)$, with the usual covariant derivative
\begin{equation}
  \label{eq_cov_dev}
\left(D_\mu\varphi\right)^a=D_\mu^{ab}\varphi^b=\partial_\mu\varphi ^a+g_0f^{abc}A_\mu^b \varphi^c
\end{equation}
for any field $\varphi$ in the adjoint representation of SU($N$), we obtain the covariant gauge condition\footnote{The gauge condition selects particular representatives  $U$ along the gauge orbit of a given field configuration $A$. In principle this can always be written as a condition in the space of field configurations, namely, in terms of $A$ alone. For instance, the Landau gauge condition can be written $\partial_\mu A_\mu=0$. In the present case, such a rewriting is difficult since the gauge transformation field $U$ genuinely appears in the gauge-fixing condition.}
\begin{equation}
  \label{eq_eq_mot}
  \left(\partial_\mu A_\mu^{U}\right)^a=\frac{ig_0}{2}\tr\left[t^a\left(U\eta^\dagger-\eta U^\dagger\right)\right].
\end{equation}

This can be used as a gauge condition for any $\eta$. Alternatively, we 
can average over $\eta$ with a given distribution ${\cal P}[\eta]$. Here, we choose a simple Gaussian 
distribution\footnote{{  We shall see below that this choice is convenient for being able to factor out the volume of the gauge group in a continuum formulation. Moreover, we mention that} more complicated distributions, involving 
either non-Gaussian or derivative terms, would lead to nonrenormalizable actions in the procedure described in Sec.~\ref{sec_averaging}.}
\begin{equation}
  \label{eq_p_eta}
  \mathcal P[\eta]={\cal N}\exp\left(-\frac {g_0^2}{4\xi_0}\int_x\tr\, \eta^\dagger\eta\right),
\end{equation}
with ${\cal N}$ a normalization factor. 

{  Equation \eqn{eq_func} is a simple generalization of the extremization functional routinely employed in lattice calculations in the Landau gauge [which is recoverd for $\eta=0$, that is by choosing $\xi_0=0$ in \Eqn{eq_p_eta}] and presents similar good properties for the purpose of numerical minimization techniques.\footnote{ Usual numerical minimization techniques require that the discretized version of the minimization functional be linear in the gauge transformation matrix $U(x)$ at each lattice point. This is the case of the standard discretization of the Landau gauge term, the first one on the right-hand side of \Eqn{eq_func} and this is obviously true for the second, $\eta$-dependent term as well; see the discussion in Appendix \ref{appsec:min}.} In this line of thought, we emphasize that a somewhat similar extremization procedure has been proposed and implemented in actual lattice calculations in Ref. \cite{Cucchieri:2009kk}.} There, the authors considered a similar functional as \eqn{eq_func}, with $i\eta$ constrained to belong to the Lie algebra of the gauge group, with the aim of enforcing a linear gauge condition.\footnote{Linear covariant gauges correspond to demanding that $\partial_\mu A_\mu^U=\phi$, with $\phi$ independent of $U$.} We see from \Eqn{eq_eq_mot} that this is only valid for gauge transformations {  close to the identity}: $U=\openone+ig_0\lambda$. Here, we do not insist on having a linear gauge fixing and \Eqn{eq_eq_mot} holds for arbitrary $U$ {  along the whole gauge orbit}. Another important difference lies in the sampling \eqn{eq_p_eta} over the matrix field $\eta$. Here, the latter is not restricted to the Lie algebra of the gauge group, which leads to a different gauge fixing ({  see below}). {  However}, we believe that the numerical implementation of Ref.~\cite{Cucchieri:2009kk} is not restricted to infinitesimal gauge transformations in principle {  and} we do not expect the different sampling on $\eta$ to be an issue {  for what concerns the question of numerical minimization}. It would thus be of great interest to investigate whether the numerical methods employed in \cite{Cucchieri:2009kk,Cucchieri:2010ku,Cucchieri:2011pp} apply to the present proposal.
 
To gain more insight on the gauge-fixing procedure described above, let us 
consider the ultraviolet regime where the standard Faddeev-Popov procedure is justified because Gribov copies are irrelevant. Setting, again, $U\to e^{ig_0\lambda}U$ in \eqn{eq_eq_mot} 
and expanding in $\lambda$, we obtain the Faddeev-Popov operator 
\begin{equation}
  \label{eq_FP_det}
 \bigg\{\partial_\mu D^{ac}_\mu[A^U]+\frac
 {g_0^2}{2}\tr\left(t^a t^cU\eta^\dagger+\eta U^\dagger t^c t^a\right)\bigg\} \delta^{(d)}(x-y)   ,
\end{equation}
where the derivatives act on the variable $x$ and where the covariant derivative, defined in \eqn{eq_cov_dev}, is to be evaluated at $A^U(x)$.
{  Introducing a Nakanishi-Lautrup field $ih$ to account for the gauge condition \eqn{eq_eq_mot} as well as ghost and antighost fields $c$ and $\cb$ to cope for the corresponding Jacobian, the Faddeev-Popov gauge-fixed action reads, for a given external field~$\eta$,
\beq
\label{eq:eff}
 S_{\rm gf}^\eta[A,c,\cb,h,U]=S_{\rm YM}[A]+S_{\rm FP}^\eta[A,c,\cb,h,U],
\eeq
with 
\begin{align}
  \label{eq_fp}
  S_{{\rm FP}}^\eta[A,c,\cb,h,U]&=\int_x\Big\{\partial_\mu\bar c^aD_\mu[A^U] c^a+ih^a\left(\partial_\mu A_\mu^U\right)^a\nn
  &\qquad\,\,\,+\frac{g_0}{2}\tr\left[\eta^\dagger R+R^\dagger\eta\right]\Big\},
\end{align}
where we introduced\footnote{Here, $ih=ih^a t^a$ is to be seen as an Hermitian matrix field and similarly for $c$ and $\bar c$.}
\beq
 R=(h-g_0\bar c c) U.
\eeq

It is important to notice here that the effective action \eqn{eq_fp} depends separately on $A$ and $U$, not only on the combination $A^U$, which makes the standard Faddeev-Popov trick of factorizing out a volume of the gauge group inapplicable. Here, the sampling \eqn{eq_p_eta} over $\eta$ is of great help since
\beq
\label{eq:trace}
 \int {\cal D}\eta{\cal P}[\eta]\,e^{-\frac{g_0}{2}\!\int_x\tr\left[\eta^\dagger R+R^\dagger\eta\right]}\propto e^{\,\xi_0\!\int_x\tr\left[R^\dagger R\right]}
\eeq
does not depend explicitly on $U$ anymore.\footnote{We note that this is not true for the sampling proposed in \cite{Cucchieri:2009kk}.} The resulting gauge-fixed action is of the form $S_{\rm YM}[A]+S_{\rm FP}[A^U,c,\cb,h]$ and one can factor out the volume of the gauge group in the standard manner. Remarkably the calculation of $\tr\left[R^\dagger R\right]$ in \eqn{eq:trace} yields, after some simple algebra,
\beq
 S_{\rm gf}[A,c,\cb,h]=S_{\rm YM}[A]+S_{\rm CFDJ}[A,c,\cb,h],
\eeq
where 
\begin{equation}
  \label{eq_action_CFDJ}
  \begin{split}
  &S_{\rm CFDJ}[A,c,\cb,h]=\int_x\bigg\{ \partial_\mu \cb^aD_\mu c^a+ih^a \partial_\mu A_\mu^a\\
  &\quad+\xi_0 \bigg[\frac {(h^a)^2}2\!-\!\frac {g_0}2 f^{abc}ih^a\cb^b c^c\!-\!\frac {g_0^2}4 \left( f^{abc}\cb^b c^c\right)^2\bigg]\bigg\}    
  \end{split}
\end{equation}
is known as the Curci-Ferrari-Delbourgo-Jarvis gauge-fixing action \cite{Curci:1976bt,Delbourgo:1981cm}. Thus, the extremization of the functional \eqn{eq_func} together with the Gaussian average \eqn{eq_p_eta} provide a nonperturbative formulation of this class of nonlinear covariant gauges. }

The CFDJ gauges have been much studied in the literature \cite{Curci:1976bt,Delbourgo:1981cm,deBoer:1995dh,Wschebor:2007vh,Tissier_08} and are known to possess various good properties. For instance, they are perturbatively 
renormalizable in four dimensions. Also, they have a nilpotent BRST symmetry and are thus unitary. 
However, they have Gribov ambiguities, just 
as the Landau gauge. This is not a problem for lattice calculations 
as one may easily select a particular copy, as done in the so-called 
minimal Landau gauge.

\section{Averaging over Gribov copies}
\label{sec_averaging}

At an analytical level, the action (\ref{eq_action_CFDJ}) suffers from the  
Neuberger zero problem \cite{Neuberger:1986vv}. The Faddeev-Popov construction 
ignores the Gribov copies, which contribute to the partition function with 
alternating signs and eventually sum up to zero. In order to cope with this issue, we follow \cite{Serreau:2012cg} and lift the degeneracy of the Gribov 
copies by means of a suitably chosen nonuniform weight.

\subsection{The general procedure}

Gribov copies correspond to the extrema $U_i\equiv U_i[A,\eta]$ of the functional ${\cal H}[A,\eta,U]$, \Eqn{eq_func}, for given $A$ and $\eta$.
For any operator $\mathcal{O}[A]$, we define the average over the Gribov copies 
of a given field configuration $A$ as\footnote{We also considered the 
following definition:
$$
  \langle\mathcal O[A]\rangle=\int \mathcal D\eta \mathcal
    P[\eta]\frac{\sum_i \mathcal O [A^{U_i}]s(i)e^{-\beta_0 
{\cal H}[A,\eta,U_i]}}{\sum_i s(i)e^{-\beta_0 {\cal H}[A,\eta,U_i]}}.
$$
Following the procedure described below, we can express this gauge fixing in 
terms of a local field theory. The latter is, however, more intricate -- it 
induces nontrivial couplings between replica, see below -- and we do not 
pursue this path further. { We emphasize that in both cases, the average over $\eta$ is performed {\em before} the one over the Yang-Mills field $A$.}
}
\begin{equation}
  \label{eq_average_G}
  \langle\mathcal O[A]\rangle=\frac{\int \mathcal D\eta \mathcal
    P[\eta]\sum_i \mathcal O [A^{U_i}]s(i)e^{-\beta_0 {\cal H}[A,\eta,U_i]}}{\int \mathcal D\eta \mathcal
    P[\eta]\sum_i s(i)e^{-\beta_0 {\cal H}[A,\eta,U_i]}},
\end{equation}
where the sums run over all Gribov copies, $s(i)$ is the sign of the functional 
determinant of the Faddeev-Popov operator \eqn{eq_FP_det} evaluated at $U={U_i}$ 
and $\beta_0$ is a free parameter which controls the lifting of degeneracy 
between Gribov copies\footnote{ We mention that the sign-weighted averages over Gribov copies proposed in \cite{Hirschfeld:1978yq} correspond to a flat weight in \eqn{eq_average_G} ($\beta_0=0$) and thus suffer from the Neuberger zero problem \cite{vonSmekal:2008en}.} according to the value of the functional ${\cal 
H}[A,\eta,U_i]$. Equation \eqn{eq_average_G} defines our gauge fixing procedure. This is inspired from the averaging procedure put forward in \cite{Tissier:2011zz} to deal with potentials with nontrivial landscapes in the context of the Random Field Ising Model.\footnote{ As was argued in \cite{Serreau:2012cg} for the case of the Landau gauge, the denominator in \eqn{eq_average_G} is a sum over real numbers -- instead of integers in the case $\beta_0=0$ -- and the set of field configurations for which it may vanish is expected to be of zero measure.}

Once the average \eqn{eq_average_G} has been performed for each individual gauge-field configuration we average over the latter with the Yang-Mills weight, hereafter denoted by an overall bar:
\begin{equation}
  \label{eq_av_A}
\overline {\mathcal O[A]}=\frac{\int\mathcal
  DA\, \mathcal O[A]e^{-S_{\rm YM}[A]}}{\int\mathcal
  DA\, e^{-S_{\rm YM}[A]}}.
\end{equation}
To summarize, our gauge-fixing procedure amounts to average first over Gribov copies and then over Yang-Mills field configurations, that is
\begin{equation}
  \label{eq_avav}
  \overline{\langle\mathcal O[A]\rangle}\,.
\end{equation}
A crucial remark is in order here; observe that gauge-invariant operators such that $\mathcal{O}_{\rm inv}[A^U]={\cal O}_{\rm inv}[A]$, are blind to the average \eqn{eq_average_G}:
$ \langle\mathcal O_{\rm inv}[A]\rangle=\mathcal O_{\rm inv} [A]$,
which guarantees that our gauge-fixing procedure does not affect physical observables. In particular, one has 
\beq
 \overline{\langle{\cal O}_{\rm inv}[A]\rangle}=\overline{{\cal O}_{\rm inv}[A]}\,.
\eeq 
It is crucial to introduce the denominator in \eqn{eq_average_G} in order for this fundamental property to hold. 

\subsection{Functional integral formulation}

The previous gauge fixing can be implemented within a
field-theoretical framework by making use of the identity
\begin{equation}
  \label{eq_gribov_FP}
  \sum_i \mathcal X [U_i]s(i)=\int {\cal D}(U,c,\bar c,h)\,\mathcal X [U] \,e^{-S_{{\rm FP}}^\eta[A^U\!\!,c,\cb,h]}    ,
\end{equation}
for any functional ${\cal X}[U]$, where $S_{{\rm FP}}^\eta$ is defined in \Eqn{eq_fp}. Here, ${\cal D}(U,c,\bar c,h)\equiv {\cal D}U{\cal D}c{\cal D}\bar c{\cal D}h$, with ${\cal D}U$ the Haar measure on the gauge group. In the following, we collect the set of fields
$U$, $c$, $\bar c$ and $h$ in a single symbol ${\cal V}$---we shall
see shortly how this can be realized explicitly in a superfield
formulation---and write $ {\cal D}(U,c,\bar c,h)={\cal DV}$. Using \eqn{eq_gribov_FP} with ${\cal X}[U]={\cal O}[A^U]\exp\{-\beta_0{\cal H}[A,\eta,U]\}$ and performing the integral over the field $\eta$ with the Gaussian measure \eqn{eq_p_eta}, we obtain
\beq
\label{eq_gf}
 \langle \mathcal O[A]\rangle=\frac{\int {\cal DV} \,O[A^U]\,e^{-S_{\rm 
CF}[A,{\cal V}]}}{\int {\cal DV}\, e^{-S_{\rm CF}[A,{\cal V}]}},
\eeq
where we denoted 
\beq 
\label{eq_actionCF_U}
 S_{\rm CF}[A,{\cal V}]\equiv S_{\rm CF}[A^U\!\!,c,\cb,h],
\eeq
 with
\beq
\label{eq_actionCF}
 S_{\rm CF}[A,c,\cb,h]=S_{\beta_0}[A,c,\cb]+S_{\rm CFDJ}[A,c,\cb,h].
\eeq
The action $S_{{\rm CFDJ}}$ is defined in \eqn{eq_action_CFDJ} and 
\begin{equation}
  \label{eq_action_W}
  S_{\beta_0}[A,c,\cb]={\beta_0}\int_x  \bigg\{\frac{1}{2}(A_\mu^a)^2+\xi_0\cb^a c^a\bigg\}.
\end{equation}
The gauge-fixing action \eqn{eq_actionCF} is a massive extension of the CFDJ 
action known as the CF action \cite{Curci:1976bt}. 
Here, the gauge-fixing parameter $\beta_0$ induces a mass for both the gluon and the ghost fields. 

We now introduce an elegant and compact superfield formulation which makes explicit some of the symmetries of the problem. First, introducing $\hat h^a=ih^a+{g_0\over2}f^{abc}\cb^b c^c$, the action \eqn{eq_actionCF} takes the ghost-antighost symmetric form
\begin{equation}
  \label{eq_av_sym}
  \begin{split}
  &S_{\text{CF}}=\int_x\bigg\{\frac {\beta_0}2 (A_\mu
  ^a)^2+\frac 12 \Big(\partial_\mu \cb^a D_\mu
  c^a+D_\mu \cb^a \partial_\mu c^a\Big)\\&+\hat h^a\partial_\mu A_\mu^a+\xi_0  \Big[\beta_0 \cb^a c^a-\frac {(\hat h^a)^2}2-\frac {g_0^2}8
\left(f^{abc}\cb^b c^c\right)^2\Big]\bigg\}.
  \end{split}
\end{equation}
The fields $U$, $c$, $\bar c$ and $\hat h$ can be put together in a matrix superfield ${\cal V}$  that depends on the Euclidean coordinate $x$ and two Grassmannian coordinates $\ts$ and $\tb$ as
\begin{equation}
  \label{eq_susy}
  \mathcal V(x,\ts,\tb)=\exp \left\{ig_0\left(\tb c+\bar
    c\ts+\tb\ts \hat h\right)\right\}U,
\end{equation}
where $\hat h$ is seen as a real field. $\mathcal V$ is a SU($N$) matrix 
field on the superspace $(x,\theta,\tb)$. It is
convenient to define an associated super gauge-field transform as
\begin{equation}
  \label{eq_super_gauge_transfo}
  A_\mu^{\mathcal V}=\mathcal V A_\mu \mathcal V^\dagger+\frac i{g_0}\mathcal
  V\partial_\mu \mathcal V^\dagger.
\end{equation}
Similarly, we introduce the pure gauge ($M=\theta,\tb$)
\beq
  \label{eq_super_pure_gauge}
   A_M^{\mathcal V}=\frac i{g_0}\mathcal V\partial_M \mathcal V^\dagger.
\eeq

The Grassmann subspace is taken to be curved with line element $ds^2=g_{MN}dNdM=2g_{\ts\tb}d\tb d\ts$, where
\beq
\begin{split}
  \label{eq_metric_grass}
  &g_{\tb\ts}=-g_{\ts\tb}=\beta_0\tb\ts+1,\\
  &g^{\tb\ts}=-g^{\ts\tb}=\beta_0\tb\ts-1.
\end{split}
\eeq
Accordingly, we define the invariant integration measure in Grassmann coordinates as  \cite{Tissier_08}
\begin{equation}
  \label{eq_note_int_grass}
\int_{\underline{\theta}}=\int d\theta d\thetab\,g^{1/2}(\theta,\thetab)  \,,
\end{equation}
where 
\begin{equation}
\label{eq_measure}
g^{1/2}(\theta,\thetab)=\beta_0\thetab\theta-1. 
\end{equation}
 Here and in 
the following, we denote the couple of Grassmann variables $(\ts,\tb)$ by 
$\underline\theta$.
It is an easy exercise to check that, in terms of the curved Grassmann space and of the fields \eqn{eq_super_gauge_transfo} and \eqn{eq_super_pure_gauge}, the Curcci-Ferrari action in \eqn{eq_gf} takes the particularly compact form of a generalized masslike term
\begin{equation}
  \label{eq_av_superfield}
 S_{\text{CF}}[A,{\cal V}]=\int_{x,\underline \theta}\tr\left\{\left(A_\mu^{\mathcal V}\right)^2 +\frac{\xi_0}2 g^{MN}A_N^{\mathcal V}A_M^{\mathcal V}\right\},
\end{equation}
which makes explicit a large group of symmetries corresponding to the isometries of the curved superspace. Equivalently, \Eqn{eq_av_superfield}
can be written as
\begin{equation}
  \label{eq_av_superfield_bis}
 S_{\text{CF}}[A,{\cal V}]\!=\!\frac{1}{g_0^2}\!\int_{x,\underline \theta}\!\tr\left\{D_\mu{\cal V}^\dagger D_\mu{\cal V} +\frac{\xi_0}2 g^{MN}\partial_N{\mathcal V}^\dagger \partial_M {\mathcal V}\right\}\!,
\end{equation}
where $D_\mu{\cal V}=\partial_\mu{\cal V}+ig_0{\cal V}A_\mu$.
This is the action of a supersymmetric nonlinear sigma model coupled to the gauge field $A_\mu$ in a gauge-invariant way. This form of the action makes explicit the invariance under the gauge transformations $A\to A^{V_R}$ and ${\cal V}\to{\cal
  V}{V}_R^{\dagger}$, with ${V_R}\equiv{V_R}(x)$ a local SU($N$) matrix.\footnote{In terms of the original fields $U$, $c$, $\cb$ and $h$, see \Eqn{eq_susy}, such transformations only  affect $U\to UV_R^\dagger$ and leave $c$, $\cb$, and $h$ invariant. Here, it is essential to recall that the action $S_{\rm CF}$ appearing in \Eqn{eq_gf} is to be evaluated at $A=A^U$. The invariance mentioned here follows from the fact that $A^U\to (A^{V_R})^{UV_R^\dagger}=A^U$.} Finally, for later purposes, it is useful to rewrite, again, \Eqn{eq_av_superfield} as
\begin{equation}
  \label{eq_av_superfield_ter}
 S_{\text{CF}}[A,{\cal V}]=\int_{x,\underline \theta}\left\{\frac 12\left(L_\mu^a-A_\mu^a\right)^2+\frac{\xi_0}4 g^{MN}L_N^aL_M^a\right\},
\end{equation}
where we introduced the vector fields
\beq
 L_\mu=\frac{i}{g_0}{\cal V}^\dagger\partial_\mu{\cal V}\quad{\rm and}\quad L_M=\frac{i}{g_0}{\cal V}^\dagger\partial_M{\cal V}
\eeq
which belong to the adjoint representation of SU($N$).

\subsection{Replicas}

The evaluation of Yang-Mills correlators or of physical quantities with our gauge-fixing procedure involves two subsequent averages, see Eqs. \eqn{eq_average_G}-\eqn{eq_avav}. The average over the Gribov copies of a given gauge-field configuration $A$ produces a complicated, highly nonlocal functional of the latter because of the nontrivial denominator in \Eqn{eq_average_G} or, equivalently, \Eqn{eq_gf}. A similar issue arises in the theory of disordered systems in statistical physics \cite{young}. Consider, for instance, an Ising model in the presence of quenched disorder, which means that the typical time scale of disorder is slow as compared to that of the Ising degrees of freedom. In that case, one first averages over statistical fluctuations of the Ising spins for a given disorder configuration and then over the possible realizations of the latter. Such two-step averages can be efficiently dealt with by using the method of replicas \cite{young}.  In the present context, the matrix fields $U$ play the role of 
the Ising spins and the gauge field $A$ of the quenched disorder. 

In its simplest version, the replica trick consists in writing formally the denominator of \Eqn{eq_gf} as
\begin{align}
  \label{eq_replica1}
  \frac{1}{\int \mathcal D {\cal V}
    \, e^{-S_{\rm CF}[A,{\cal V}]}}&=\lim_{n\to 
0}\left(\int \mathcal D {\cal V}
    \, e^{-S_{\rm CF}[A,{\cal V}]}\right)^{n-1}\nn
    &=\lim_{n\to 
0}\int\prod_{k=1}^{n-1}\left( \mathcal D {\cal V}_k
    \, e^{-S_{\text{CF}}[A,\mathcal V_k]}\right).
\end{align}
Here and below, the limit is to be understood as the value of the (analytically
continued) function of $n$ on the right-hand side when $n\to0$.
The average over the disorder field $A$ can then be formally written as 
\begin{equation}
  \label{eq_average2}
  \overline{\langle{\cal O}[A]\rangle}=\lim_{n\to 0}\frac{\int\mathcal D A\left(\prod_{k=1}^n \mathcal D {\cal V}_k\right)\,{\cal O}[A^{U_1}]\, e^{-S[A,\{{\cal V}\}]}}{\int{\cal D}A\,e^{-S_{\rm YM}[A]}}\,,
\end{equation}
where
\beq
\label{eq:action1}
 S[A,\{{\cal V}\}]=S_{\rm YM}[A]+\sum_{k=1}^{n}S_{\rm CF}[A,{\cal V}_k]\,.
\eeq
Finally, using \Eqn{eq_average2} with ${\cal O}[A]=1$, we obtain the more convenient expression 
\begin{equation}
  \label{eq_average2bis}
  \overline{\langle{\cal O}[A]\rangle}=\lim_{n\to 0}\frac{\int\mathcal D A\left(\prod_{k=1}^n \mathcal D {\cal V}_k\right)\,{\cal O}[A^{U_1}]\, e^{-S[A,\{{\cal V}\}]}}{\int\mathcal D A\left(\prod_{k=1}^n \mathcal D {\cal V}_k\right)\, e^{-S[A,\{{\cal V}\}]}}\,.
\end{equation}
Here, the choice of the replica $k=1$ is arbitrary because of the obvious symmetry between replicas.

It may be necessary, e.g.,  for analytic approaches, to explicitly factor out the volume of the gauge
group $\int{\cal D}U$. This can be done by performing the change of variables 
$A\to A^{U_1}$ and $U_{k}\to U_k U_1^{-1}$,  $\forall\,\, k>1$ in \eqn{eq_average2bis}. Renaming $(c_1,\cb_1,h_1)\to (c,\cb,h)$, we get
\begin{equation}
  \label{eq_average2ter}
  \overline{\langle{\cal O}[A]\rangle}=\lim_{n\to 0}\frac{\int\mathcal D (A,c,\bar c,h,\{{\cal V}\})\,{\cal O}[A]\, e^{-S[A,c,\bar c,h,\{{\cal V}\}]}}{\int\mathcal D (A,c,\bar c,h,\{{\cal V}\})\, e^{-S[A,c,\bar c,h,\{{\cal V}\}]}}\,,
\end{equation}
with $\mathcal D (A,c,\bar c,h,\{{\cal V}\})\equiv\mathcal D (A,c,\bar c,h)\left(\prod_{k=2}^n \mathcal D {\cal V}_k\right)$ and
\begin{equation}
  \label{eq_action2}
  \begin{split}
  S[A,c,\bar c,h,\{{\cal V}\}]&=S_{{\rm YM}}[A]+S_{{\rm CF}}[A,c,\cb,h]\\
  &+\sum_{k=2}^n S_{\rm CF}[A,\mathcal V_k],
      \end{split}
\end{equation}
where we used the notation \eqn{eq_actionCF_U} in the last line.
Thus, we see that \Eqn{eq:action1} describes a collection of $n$ gauged supersymmetric nonlinear
sigma models coupled to the Yang-Mills field $A$. It is invariant under the local right color rotations $A\to A^{V_R}$ and ${\cal V}_k\to{\cal
  V}_kV_R^\dagger$, $\forall k=1,\ldots,n$.  This symmetry gets
explicitly broken after one replica is singled out to extract the
volume of the gauge group. The action \eqn{eq_action2} describes $n-1$ gauged supersymmetric nonlinear sigma
models coupled to a gauge-fixed Yang-Mills field with gauge-fixing action $S_{\rm CF}[A,c,\cb,h]$. As discussed below the action \eqn{eq_action2} possesses a BRST symmetry as a remnant of the original gauge symmetry. 

\section{Renormalizability}
\label{sec_renorm}

\subsection{Symmetries}
\label{sec:symmetries}

We now prove the perturbative renormalizability of the action
\eqn{eq_action2} in $d=4$. This is nontrivial given the 
presence of nonlinear sigma models, which
are, in general, renormalizable in $d=2$.  Our proof follows standard
arguments \cite{WeinbergBook,Zinn-Justin-book} and consists in identifying all local
terms of mass dimension less than or equal to 4,\footnote{This relies on
  Weinberg's theorem and assumes, in particular, that the free
  propagators decrease sufficiently fast at large momentum. We show in
  the next section that all free propagators decrease at least as fast as $1/p^2$ at
  large $p$, which is a sufficient condition.} compatible with the symmetries of the theory in the effective action
$\Gamma$.

Let us first list the symmetries of the action \eqn{eq_action2} that
are realized linearly. Apart from the global SU($N$) color
symmetry and the isometries of the Euclidean space $\mathbb{R}^4$, there are the
net ghost number conservation ($c\to e^{i\epsilon}c$, $\cb\to
e^{-i\epsilon} \cb$) and the isometries of the curved Grassmann
space. The latter only impact the superfields: $\mathcal
V_k\to\mathcal
V_k+X^{M}\partial_{M}\mathcal V_k$, with $M=\theta,\tb$,
where $X^{M}$ is one of the five independent Killing
vectors on the Grassmann space \cite{Tissier_08}. At the level of the
effective action, these symmetries simply imply that terms involving
Grassmann variables should be written in a covariant way: integrals
always come with the proper integration measure, see
\eqn{eq_note_int_grass}, and derivatives are contracted with proper
tensors \cite{Tissier_08}. An important remark to be made is that
these transformations apply to each individual replica superfield
$\mathcal V_k$, independently of the others. This implies that each
such superfield comes with its own set of Grassmann
variables. There is also a discrete symmetry under the permutation of the
replicas: ${\cal V}_k\leftrightarrow{\cal V}_l$ for $k,l=2,\ldots,n$.
 
These  linear  transformations are also symmetries of the effective action
$\Gamma$ and directly constrain the possible divergent terms. We shall also
exploit the fact that the choice of the replica $k=1$
singled out in \eqn{eq_average2ter}-\eqn{eq_action2}, being arbitrary,
the divergences associated with the fields $c$, $\cb$, and $h$ are the
same as those associated with $c_k$, $\cb_k$, and $h_k$ for
$k\ge2$.\footnote{For instance, upon the change of variables $A\to
  A^{U_2}$, $U_k\to U_kU_2^{-1}$ for $k>2$, $U_2\to U_2^{-1}$ and
  $c\leftrightarrow c_2$, $\cb \leftrightarrow \cb_2$ and
  $h\leftrightarrow h_2$, one gets that it is now the replica $k=2$
  which is singled out.}
  
The action \eqn{eq_action2} also admits nonlinear symmetries. One is a
BRST-like symmetry, corresponding to the infinitesimal transformation
\begin{equation}
  \label{eq_brst}
  \begin{split}
    sA_\mu^a&=\partial_\mu c^a+g_0f^{abc}A_\mu^bc^c,\\
    sc^a&=-\frac {g_0}2f^{abc}c^bc^c,\\
    s\cb^a&=ih^a,\\
     sih^a&=\beta_0 c^a
  \end{split}
\end{equation}
and 
\beq
\label{eq_BRST_V}
 s\mathcal V_k=-ig_0\mathcal V_k c\,, \quad k=2,\ldots,n\,.
\eeq
In the sector $(A,c,{\cal V}_k)$ this simply corresponds to a gauge 
transformation with Grassmann parameters 
$c^a$; see the discussion below \Eqn{eq_action2}. 
For the analysis to follow, it proves convenient to employ the linear parametrization of the SU($N$)
superfield $\mathcal V_k$
\begin{equation}
  \label{eq_param_SUN}
  \mathcal V_k=(a^0_k+ib^0_k)\openone+i(a^a_k+ib^a_k)t^a
\end{equation}
with an implicit sum over the $N^2-1$ color indices. Here, we choose the $a^a_k$ as the $N^2-1$ unconstrained superfields. The fields $a^0_k$, $b^0_k$, and $b^a_k$ are functions of $a^a_k$, determined by the constraint that $\mathcal
V_k\in$ SU($N$). In practice, we will not need their explicit expressions. The BRST transformation \eqn{eq_BRST_V} of the basic field $a_k^a$ reads
\beq
 sa_k^a=g_0\left(-a_k^0c^a+\frac {1}2f^{abc}a_k^bc^c+\frac {1}2d^{abc}b_k^bc^c\right)
\eeq
and those of the constrained fields are
\begin{equation}
\label{eq_brst2}
\begin{split}
 sa_k^0&=\frac{g_0}{2N}a_k^bc^b\,,\qquad sb_k^0=\frac{g_0}{2N}b_k^bc^b\,,\\
 sb_k^a&=g_0\left(-b_k^0c^a+\frac {1}2f^{abc}b_k^bc^c-\frac {1}2d^{abc}a_k^bc^c\right)  .
\end{split}
\end{equation}

It is easy to check that  $s^2=0$ in 
the sector $(A,c,{\cal V}_k)$. The full BRST transformation is, however, 
not nilpotent: $s^2=\beta_0 t$, where $t$ is another nonlinear symmetry of the 
problem \cite{Delduc:1989uc},
whose action on the primary fields is
\begin{equation}
  \label{eq_t}
  \begin{split}
    t\cb^a&=c^a,\\
    t ih^a&=-\frac{g_0}{2}f^{abc}c^bc^c
  \end{split}
\end{equation}
and $tA_\mu^a=tc^a=t\mathcal V_k=0$.

Finally, there is a third family of nonlinearly realized symmetries, one for each replica, which corresponds to global left color rotations of the nonlinear sigma model fields ${\cal V}_k\to V_{L,k}{\cal V}_k$ with $V_{L,k}\in$ SU($N$). These symmetries are nonlinear because ${\cal
  V}_k$ is a constrained superfield. Each replica superfield can be
transformed independently of the others and there are thus
$(N^2-1)\times (n-1)$ generators. The infinitesimal
transformations are 
\beq
 \delta_k^a\mathcal V_l=i\delta_{kl}t^a\mathcal V_l.
\eeq
In terms of the representation \eqn{eq_param_SUN}, the transformation of the basic field $a_k^a$ reads
\beq
 \delta_k^a a_l^b=\delta_{kl}\left(\delta^{ab} a^0_k+\frac 12f^{abc}a^c_k-\frac 12d^{abc}b^c_k\right)
\eeq
and those of the constrained fields are
\beq
\label{eq_nl_color22}
\begin{split}
   \delta_k^a a_l^0&=-\delta_{kl}\frac {a^a_k}{2N}\,,\qquad\delta_k^a b_l^0=-\delta_{kl}\frac {b^a_k}{2N},\\
\delta_k^a b_l^b&=\delta_{kl}\left(\delta^{ab} b^0_k+\frac 12f^{abc}b^c_k+\frac 12d^{abc}a^c_k\right).
\end{split}
\eeq
We mention that the generators of the nonlinear symmetries considered above induce a
closed (super)algebra:
\beq
\label{eq_algebra}
\begin{split}
 &\{s,s\}=2\beta_0t\,,\\
 &[\delta_k^a,\delta_l^b]=i\delta_{kl}f^{abc}\delta_k^c\,,\\
 & [\delta_k^a,s]=[\delta_k^a,t]=[s,t]=0  \,.  
\end{split}
\eeq

We wish to derive Slavnov-Taylor identities associated with the nonlinear
symmetries described above in the form of Zinn-Justin equations \cite{WeinbergBook,Zinn-Justin-book}. To this aim, we introduce
(super)sources coupled to both the (super)fields and their variations
under $s$ and $\delta^a_k$. We define\footnote{Note that the variations of the (super)fields under $s^2$, $s\delta^a_k=\delta^a_k s$, and
  $\delta^a_k\delta^b_l$ can be fully expressed in terms of either the
  (super)fields themselves or their variations under $s$ or
  $\delta^a_k$. Therefore, they do not require independent
  (super)sources.}
\bea
  \label{eq_sources}
  S_1&=&\int_x\Big\{\!J_\mu^a A_\mu^a\!+\!\bar \eta^a
  c^a\!+\!\cb^a \eta^a\!+\!ih^aM^a\!+\!\bar K_\mu^a sA_\mu^a\!+\!\bar L^asc^a\!\Big\} \nn
  &+&\sum_{k=2}^n\int_{x,\underline{\theta}}\Big\{P^0_ka^0_k+P^a_ka^a_k+R^0_kb^0_k+R^a_kb^a_k\nn
  &&\qquad\quad+\bar Q^0_ksa^0_k+\bar Q^a_ksa_k^a+\bar T^0_ksb^0_k+\bar T^a_ksb_k^a\Big\}
\eea
and consider the Legendre transform $\Gamma$ of the functional
$W=\ln\int\mathcal D(A,c,\cb,h,\{{\cal V}\}) \,e ^{-S+S_1}$ with respect to the sources
$J^a_\mu$, $\eta^a$, $\bar \eta^a$, $M^a$ and $P_k^a$.  It is a straightforward procedure to derive the
desired identities \cite{WeinbergBook,Zinn-Justin-book}. 

Following \cite{Zinn-Justin-book}, it proves convenient to introduce 
a generalized transformation $\tilde s$ as
\begin{equation}
\label{eq_def_sr}
  \tilde s=\sum_{\varphi}\int_x(\tilde s \varphi)\frac{\delta\ }{\delta \varphi}+\sum_{k=2}^n \int_{x,\underline\theta} (\tilde s a_{k}^a)\frac{\delta_\theta\ }{\delta a_k^a}
\end{equation}
where the sum runs over the fields $\varphi=A, c, \cb,ih$. Here, we 
introduced the covariant functional derivative $\delta_\theta \Gamma/\delta 
\phi=(g^{-1/2})\delta
\Gamma/\delta \phi$, with any superfield $\phi$, where the metric factor $g$ is 
defined in \eqn{eq_note_int_grass}. This accounts for the curved Grassmann
directions. The variations of the fields are defined as
\beq
\label{eq_BRSTr}
 \tilde s A_\mu^a=-\frac{\delta\Gamma}{\delta \bar K_\mu^a}\,,\quad
 \tilde s c^a=-\frac{\delta\Gamma}{\delta \bar L^a}\,,\quad
 \tilde s a_k^a=-\frac{\delta_\theta\Gamma}{\delta \bar Q_k^a}
\eeq
and 
\beq
  \tilde s \bar c^a=ih^a\,,\quad \tilde s ih^a=\beta_0c^a.
\eeq
In analogy with \Eqn{eq_def_sr}, we introduce the 
transformations $\tilde t$ and $\tilde \delta_k^a$ defined by their action on 
the primary fields
\beq
 \tilde t\bar c^a=c^a\,,\quad \tilde tih^a=-\frac{\delta\Gamma}{\delta \bar L^a}
\eeq
and
\begin{equation}
\label{eq_delta_r}
  \tilde\delta_{k}^aa_l^b=  \delta_{kl}\left(-\delta^{ab} \frac{\delta_\theta\Gamma}{\delta P_k^0}+\frac 12f^{abc}a^c_{k}+\frac 12d^{abc}\frac{\delta_\theta\Gamma}{\delta R_k^a}\right)
\end{equation}
 with all other variations being zero. With these notations the relevant 
symmetry identities read

\beq
\label{eq_ST_BRS}
 \tilde s\Gamma=\sum_{k=2}^n\int_{x,\underline{\theta}}\bigg\{P_k^0\frac{\delta_ \theta
      \Gamma}{\delta \bar Q_k^0}+R_k^0\frac{\delta_ \theta
      \Gamma}{\delta \bar T_k^0}+R_k^a\frac{\delta_ \theta
      \Gamma}{\delta \bar T_k^a}\bigg\},
\eeq
\beq
\label{eq_ST_t}
 \tilde t\Gamma=0
\eeq
and
\begin{widetext}
\beq
\label{eq_ST_color}
 \begin{split}
  \tilde\delta_{k}^a\Gamma= 
\int_{x,\underline\theta}&\bigg\{\frac
1{2N}\left(P^0_ka^a_k-R^0_k\frac{\delta_\theta \Gamma}{\delta R^a_k}-\bar
  Q^0_k\frac{\delta_\theta \Gamma}{\delta \bar Q^a_k}-\bar T^0_k\frac{\delta_\theta
    \Gamma}{\delta \bar T^a_k}\right)+\left(R^a_k\frac{\delta_\theta \Gamma}{\delta R^0_k}+\bar
  Q^a_k\frac{\delta_\theta \Gamma}{\delta \bar Q^0_k}+\bar T^a_k\frac{\delta_\theta
    \Gamma}{\delta \bar T^0_k}\right)  \\
    &\,+\frac{f^{abc}}{2}\left(R^b_k\frac{\delta_\theta
    \Gamma}{\delta R^c_k}+\bar Q^b_k\frac{\delta_\theta \Gamma}{\delta \bar
    Q^c_k}+\bar T^b_k\frac{\delta_\theta \Gamma}{\delta \bar T_k^c} \right) 
-\frac{d^{abc}}{2}\left(R^b_ka^c_k+\bar Q^b_k\frac{\delta_\theta \Gamma}{\delta \bar
    T^c_k}-\bar T^b_k\frac{\delta_\theta \Gamma}{\delta \bar Q^c_k} \right) \bigg\}.
  \end{split}
\eeq
\end{widetext}

\subsection{Constraining ultraviolet divergences}
\label{sec_constraining}

The proof of renormalizability follows standard lines 
\cite{WeinbergBook,Zinn-Justin-book}. It eventually boils down to constraining 
the form of the divergent part $\Gamma^{\rm div}$ of the effective action 
through the Zinn-Justin equations derived in the previous subsection. Standard 
power counting arguments imply that $\Gamma^{\rm div}=\int d^4x\,{\cal L}^{\rm 
div}(x)$ with ${\cal L}^{\rm div}$ the most general local Lagrangian density 
including operators of mass dimension lower than or equal to 4, compatible with the 
symmetries of the problem. Here, one must take into account the fact that the 
Grassmann integration measure has mass dimension two:\footnote{This can be seen 
as follows. The metric of the Grassmann subspace \eqn{eq_metric_grass} must be 
dimensionless. It follows that $[\ts]=[\tb]=[\beta_0^{-1/2}]=-1$. The 
identity $\int d\ts\, \ts=\int d\tb\,\tb=1$ then implies that 
$[d\ts]=[d\tb]=1$.} $[d\theta d\thetab]=2$. This implies that a contribution to 
${\cal L}^{\rm div}$ of
the form $\int_{\underline \theta}\mathcal L_2(x,\underline\theta)$ is such that  the 
Lagrangian density $\mathcal L_2$ is of mass dimension $2$; a contribution of the
form $\int_{\underline \theta,\underline \theta'}\mathcal 
L_3(x,\underline\theta,\underline\theta')$ is such that $\mathcal L_3$ is of 
mass dimension
$0$, etc.

The constraints from the linear symmetries listed below are trivially 
implemented. In order to write the most general local Lagrangian ${\cal L}^{\rm 
div}$ consistent with power counting and those symmetries, it is convenient to 
recall the dimension and ghost numbers of the building blocks of the action. 
These are resumed in Table~\ref{tab_dim}. Finally, we recall that each replica 
comes with its own set of Grassmann variables and thus with its own set of 
isometries.  We can thus make a joint expansion in the number of free replica 
indices and Grassmann integrals. 
\begin{table}[h!]
  \centering
  \begin{tabular}{|l|c|c|c|c|c|c|c|c|c|c|c|c|c|c|c|c|c|}
\hline
&$A$&$c$&$\cb$&$ih$&$a$&$\bar K$&$\bar L$&$ P$&$ R$&$\bar Q$&$\bar T$&$\theta$&$\partial_\theta$&$d\theta$&$\tb$&$\partial_\tb$&$d\tb$\\
\hline
dim.&1&1&1&2&0&2&2&2&2&1&1&-1&1&1&-1&1&1\\
\hline
ghost nb. &0&1&-1&0&0&-1&-2&0&0&-1&-1&1&-1&-1&-1&1&1\\
\hline
  \end{tabular}
\caption{Mass dimension and ghost number of the fields, sources,
and Grassmann coordinates.\label{tab_dim}}
\end{table}
By inspection, we see that the divergent terms are at most linear in
the sources. By analogy with (\ref{eq_sources}), we write
\begin{equation}
  \label{eqgam0gam1}
    \Gamma_{\rm div}=\Gamma_0-\Gamma_1,
\end{equation}
where $\Gamma_0$ is independent of the sources and 
\begin{equation}
  \label{eq_ansatz_gamma1}
  \begin{split}    
\Gamma_1&=\int_x\Big\{\bar K_\mu^a \tilde sA_\mu^a+\bar L^a\tilde sc^a\Big\} \\ 
&+\sum_{k=2}^n\int_{x,\underline{\theta}}\Big\{P^0_{k}\tilde a^0_{k}+R^0_{k}\tilde b^0_{k}+R^a_k\tilde b^a_{k}\\ 
&\quad\quad+\bar Q^0_kX^0_{k}+\bar Q^a_k\tilde s a_{k}^a+\bar T^0_kY^0_{k}+\bar T^a_kY_{k}^a\Big\}.
  \end{split}
\end{equation}  
Here, the unknown functions $\tilde sA_\mu^a$, $\tilde sc^a$, and $\tilde s 
a_k^a$ are defined in \Eqn{eq_BRSTr} with $\Gamma\to\Gamma_{\rm div}$. Notice 
that the functions $\tilde a_{k}^0$, $\tilde b_{k}^0$, and $\tilde b_{k}^a$ are 
of dimension zero and can only depend on the superfields $a_l^b$.
When restricted to the divergent part of the effective action $\Gamma_{\rm div}$, the variation \eqn{eq_delta_r} thus reads
\begin{equation}
\label{eq_delta_r_div}
  \tilde\delta_{k}^aa_l^b=  \delta_{kl}\left(\delta^{ab} \tilde a^0_{k}+\frac 
12f^{abc}a^c_{k}-\frac 12d^{abc}\tilde b^c_{k}\right).
\end{equation}
Inserting \Eqn{eq_ansatz_gamma1} in the symmetry identity \eqn{eq_ST_BRS} and
extracting the terms linear in $P_k^0$, $R_k^0$, and $R_k^a$, we find that
\begin{equation}
  \label{eq_}
  X_k^0=\tilde s \tilde a_{k}^0\,,\qquad Y_k^0=\tilde s \tilde b_{k}^0\,,\qquad Y_k^a=\tilde s \tilde b_{k}^a\,,
\end{equation}
with the transformation $\tilde s$ defined in \eqn{eq_def_sr}.
Similarly, extracting the terms linear in the remaining sources in
Eq.~(\ref{eq_ST_BRS}) as well as the terms linear in the sources in
Eqs.~(\ref{eq_ST_t}) and (\ref{eq_ST_color}), we conclude that the
renormalized transformations $\tilde s$, $\tilde t$, and $\tilde\delta_k^a$ satisfy the same algebra as the bare
ones, Eq.~(\ref{eq_algebra}), with the bare parameter $\beta_0$
appearing explicitly. 

In order to find the most general form for the transformations
$\tilde s$ and $\tilde\delta_{k}^a$, it proves convenient to group the fields 
$a_k^a$ and the unknown functions $\tilde a_k^0$, $\tilde b_k^0$, and 
$\tilde b_k^a$ in the matrix
\begin{equation}
  \label{eq_param_SUN_2}
  \tilde {\mathcal V}_{k}=(\tilde a^0_{k}+i\tilde b^0_{k})\openone+i(a^a_k+i\tilde b^a_{k})t^a.
\end{equation}
The operator $\tilde s$ is of dimension one and
has a ghost number one. By inspection, we find the most general form
of the renormalized BRST variations of the fields to be
\begin{equation}
  \label{eq_brst_r}
  \begin{split}
    \tilde sA_\mu^a&=\kappa_1\partial_\mu c^a+\tilde g f^{abc}A_\mu^bc^c\,,\\
\tilde sc^a&=-\frac {\tilde g}2f^{abc}c^bc^c\,, \\
\tilde s\tilde{\mathcal V}_{k}&=-i\tilde g\tilde{\mathcal V}_{k} c\,, \quad k=2,\ldots,n\,.
  \end{split}
\end{equation}
Similarly we get, for the most general form of the transformation $\tilde\delta_{k}$,
\begin{equation}
\label{eq_rotationV_r}
\tilde\delta_{k}^a\tilde{\mathcal V}_{l}=i\delta_{kl}t^a\tilde {\mathcal V}_{l}  .
\end{equation}
which shows that $\tilde {\mathcal V}_k$ transforms under a linear 
representation of SU$(N)$. It follows that 
\beq
\label{eq_oneZ}
 \tilde {\cal V}_{k}^\dagger\tilde{\cal V}_{k}=Z\openone,
\eeq
with $Z$ a (possibly divergent) constant.

Finally, there remains to determine the source-independent term $\Gamma_0$, 
which satisfies
\begin{equation}
  \tilde s \Gamma_0=\tilde t\Gamma_0=\tilde\delta_{k}^a\Gamma_0=0.
\end{equation}
Using the fact that, by power counting, there can be at most two set of Grassmann variables, we parametrize the solution as 
\begin{equation}
  \label{eq_ansatz_gamma0}
  \begin{split}    
\Gamma_0&=\int_x\mathcal L_1(A,c,\cb,h)  +\sum_{k=2}^n\int_{x,\underline{\theta}}\mathcal L_2(A,c,\cb,h,a_k(\underline\theta))\\&+\sum_{k,k'=2}^n\int_{x,\underline{\theta},\underline{\theta}'}\mathcal L_3(a_k(\underline\theta),a_{k'}(\underline\theta')).    
  \end{split}
\end{equation}  
Power counting implies that ${\cal L}_3$ is of mass dimension zero. Therefore, it cannot involve the fields $A$, $c$, $\cb$, or $h$. Similarly,
it cannot involve any derivatives $\partial_\mu$ or $\partial_M$. It is thus a potential term for the
superfields $a_k$ and $a_{k'}$ (or equivalently $\tilde{\mathcal V}_k$ and 
$\tilde{\mathcal V}_{k'}$). The only possible such term compatible with the 
symmetry \eqn{eq_rotationV_r} is a function of $\tilde {\cal 
V}_{k}^\dagger\tilde{\cal V}_{k}$ and $\tilde {\cal V}_{k'}^\dagger\tilde{\cal 
V}_{k'}$, which is trivial due to \eqn{eq_oneZ} so that $\mathcal L_3=0$.

Notice that, in the sector ($A, c, \tilde{\cal V}_{k}$), the transformation $\tilde s$ is, up to a multiplicative factor $\kappa_1$, a (left) gauge transformation with Grassmannian gauge parameter $c^a$ and effective coupling constant $\tilde g/\kappa_1$. A trivial solution to $\tilde s{\cal L}_1=0$ is thus a Yang-Mills-like term with an appropriate field-strength tensor, see below. It is easy to check that apart from this term, the combinations 
\bea
 X&=&\frac{\beta_0}{2\kappa_1}\left(A_\mu^a\right)^2-\tilde s\left( A_\mu^a\partial_\mu\cb^a\right),\\
 Y&=&\beta_0\cb^ac^a-\tilde s\left[\cb^a\left(ih^a+\frac{\tilde g}2f^{abc}\cb^b 
c^c\right)\right]
\eea 
are the only independent solutions to $\tilde s{\cal L}_1=0$ with the correct dimension, symmetries, and ghost number. Thus
\beq
 {\cal L}_1=\frac {Z_1}4(\tilde F_{\mu\nu}^a)^2+\kappa_2X+\frac{\kappa_3}2Y,
\eeq
with 
\beq
 \tilde F_{\mu\nu}^a=\partial_\mu A_\nu^a-\partial_\nu A_\mu^a +\frac{\tilde g}{\kappa_1}f^{abc}A_\mu^b A_\nu^c.
\eeq
Explicitly, one has
\begin{equation}
  \label{eq_L1}
  \begin{split}
  &\mathcal L_1\!=\!\frac {Z_1}4(\tilde F_{\mu\nu}^a)^2\!+\!\kappa_2\!\left\{\frac{\beta_0}{2\kappa_1}(A_\mu^a)^2-iA_\mu^a\partial_\mu h^a+\partial_\mu\cb^a\tilde sA_\mu^a\right\}\\
  &+\kappa_3\!\left\{\beta_0\cb^a c^a+\frac{(h^a)^2}2-\frac {\tilde g}2 f^{abc}ih^a\cb^bc^c-\frac{\tilde g^2}4(f^{abc}\cb^bc^c)^2\right\}\!,
  \end{split}
\end{equation}
with $\tilde sA_\mu^a$ given in \eqn{eq_brst_r}. This is trivially invariant under $\tilde t$ and $\tilde\delta_{k}^a$.

Let us now consider the nonlinear sigma model sector ${\cal L}_2$. The 
constraint $\tilde\delta_{k}^a{\cal L}_2=0$ is trivially accounted for by using 
the SU$_R(N)$ invariants $\tilde{\cal 
V}_{k}^\dagger\partial\dots\partial\tilde{\cal V}_{k}$, with an arbitrary number 
of bosonic and Grassmannian derivatives, as building blocks. The term with no 
derivatives is trivial due to \eqn{eq_oneZ}. The 
isometries of the embedding superspace and the fact that ${\cal L}_2$ can only 
contain local terms of mass dimension lower than two restricts the set of 
possible invariants to ($M=\theta,\bar\theta$)
\beq
 \tilde L_{k,\mu}=\frac {i}{\tilde g} \tilde{\cal V}_{k}^\dagger  \partial_\mu\tilde {\cal V}_{k}\quad {\rm and}\quad \tilde L_{k,M}=\frac {i}{\tilde g} \tilde{\cal V}_{k}^\dagger  \partial_M\tilde{\cal V}_{k}\,.
\eeq
Both $\tilde L_{k,\mu}$ and $\tilde L_{k,M}$ have mass dimension one. Their ghost numbers are $0$ for $\tilde L_{k,\mu}$, $1$ for $\tilde L_{k,\bar\theta}$ and $-1$ for $\tilde L_{k,\theta}$.  The variation of $\tilde L_{k,\mu}$ under $\tilde s$ is
\beq
 \tilde s \tilde L_{k,\mu}^a=Z\partial_\mu c^a+\tilde gf^{abc}\tilde L_{k,\mu}^bc^c
\eeq
It follows that $\tilde L_{k,\mu}/Z-A_\mu/\kappa_1$ transforms covariantly
\beq
 \tilde s\left(\tilde L_{k,\mu}^a-\frac{Z}{\kappa_1}A_\mu^a\right)=\tilde gf^{abc}\left(\tilde L_{k,\mu}^b-\frac{Z}{\kappa_1}A_\mu^b\right)c^c.
\eeq
Similarly, $\tilde L_{k,M}$ transform covariantly:
\beq
 \tilde s\tilde L_{k,M}^a=-\tilde gf^{abc}\tilde L_{k,M}^bc^c.
\eeq
The most general dimension two Lagrangian ${\cal L}_2$ satisfying $\tilde s{\cal L}_2=0$ is thus
\begin{equation}
  \label{eq_L2}
  \mathcal L_2=\frac{Z_2}2\left(\tilde L_{k,\mu}^a-\frac{Z}{\kappa_1}A_\mu^a\right)^2+\frac{Z_3}4g^{MN}\tilde L_{k,N}^a\tilde L_{k,M}^a
\end{equation}

We see that the most general divergent part compatible with the symmetries has 
the same form as the bare Lagrangian. This demonstrates the (multiplicative) 
renormalizability of the present theory. So far we have eight independent 
renormalization constants $\kappa_{1,2,3}$, $Z_{1,2,3}$, $Z$, and $\tilde g$. As 
described in Appendix \ref{appsec:replica}, the original symmetry between the 
replicas $k=1$ and $k\ge2$, mentioned in Sec.~\ref{sec:symmetries}, leads to 
the relations
\beq
\label{eq_symreprel}
 Z_2Z^2=\kappa_1\kappa_2\quad{\rm and}\quad Z_3Z^2=\kappa_3
\eeq
which reduce the number of independent renormalization constants to six.
In particular, it follows that all replicas contribute a mass term for the gauge field
\beq
 \frac{Z_2Z^2}{2\kappa_1^2}\int_{\underline\theta}(A_\mu^a)^2=\frac{\beta_0\kappa_2}{2\kappa_1}(A_\mu^a)^2,
\eeq
identical to the one in \eqn{eq_L1}. The total $A^2$ contribution is thus proportional to $n$, as expected from the replica symmetry. 

\subsection{Relation with perturbation theory}
\label{sec_renorm_pert}

To make link with perturbation theory, we introduce the usual renormalized fields and constants as
\beq
\label{eq_renfields}
 A= \sqrt{Z_A} A_r\,,\,\, c=\sqrt{Z_c} c_r\,,\,\,\cb=\sqrt{Z_c}\cb_r\,,\,\, ih=\sqrt{Z_h}i h_r
\eeq
and
\beq
\label{eq_renconst}
 \beta_0=Z_\beta\beta\,,\quad\xi_0=Z_\xi \xi\,,\quad g_0=Z_g g.
\eeq
It is useful to also introduce rescaled Grassmann variables $\ts_r$ and $\tb_r$ 
such that the measure \eqn{eq_measure} reads 
$\beta_0\bar\theta\theta-1=\beta\bartheta_r\theta_r-1$. We thus define
\beq
\label{eq_Grass_renorm}
\begin{split}
 &\theta= Z_\beta^{-1/2} \theta_r\,,\quad\partial_\theta=Z_\beta^{1/2}\partial_{\theta_r}\,,\quad d\theta= Z_\beta^{1/2}d\theta_r\,,\\
 &\bar\theta= Z_\beta^{-1/2}\bar\theta_r\,,\quad\partial_{\bar\theta}= Z_\beta^{1/2}\partial_{\bar\theta_r}\,,\quad d\bar\theta= Z_\beta^{1/2}d\bar\theta_r\,.
\end{split}
\eeq
Accordingly we introduce a renormalized metric as
\beq
\label{eq_renorm_metric}
 g_r^{M_rN_r}(\underline{\theta_r})=g^{MN}(\underline\theta),
\eeq
with $M_r,N_r=\ts_r,\tb_r$. In particular, this implies
\beq
\label{eq_ren_Grass_int}
 \int_{\underline\theta}=Z_\beta\int_{\underline{\theta_r}},
\eeq
where $\int_{\underline{\theta_r}}=\int d\theta_r d\bar\theta_r(\beta\bar\theta_r\theta_r-1)$.
Finally, parametrizing the bare nonlinear model superfields as
\beq
\label{eq_param}
 {\cal V}_{k}=\exp\{ig_0\Lambda_{k}\},
\eeq
we define the corresponding renormalized superfields as
\beq
\label{eq_renormL}
 \Lambda_{k}=\sqrt{\frac{Z_\Lambda}{Z_\beta}}\,\Lambda_{r,k}.
\eeq
The replica symmetry imply that $Z_\Lambda$ is the same for all $k\ge2$. Here, 
we extracted a factor $\sqrt{Z_\beta}$ in such a way that  the kinetic term of 
the fields $\Lambda_{r,k}^a$ is normalized as 
\beq
 \frac12 \int_{\underline\theta}\left(\partial_\mu\Lambda_k^a\right)^2=\frac{Z_\Lambda}2\int_{\underline{\theta_r}}\left(\partial_\mu\Lambda_{r,k}^a\right)^2.
\eeq
For later use, we also mention the identities
\beq
 \int_{\underline\theta}A_\mu^a\partial_\mu\Lambda_k^a=\sqrt{Z_\beta Z_\Lambda Z_A}\int_{\underline{\theta_r}}A_{r,\mu}^a\partial_\mu\Lambda_{r,k}^a
\eeq
and
\bea
&& \frac{\xi_0}{4}\int_{\underline\theta}g^{MN}\partial_N \Lambda_{k}^a\partial_M\Lambda_k^a=\frac{\xi_0}{4}\int_{\underline\theta}\Lambda_k^a\square_{\underline\theta}\Lambda_k^a\nn
 &&\hspace{1.6cm}=Z_\beta Z_\Lambda Z_\xi\,\frac{\xi}{4}\int_{\underline{\theta_r}}\Lambda_{r,k}^a\square_{\underline{\theta_r}}\Lambda_{r,k}^a,
\eea
where 
 \begin{align}
\square_{\underline\theta}&=\frac{1}{\sqrt{g}}\partial_M\sqrt{g}g^{MN}
\partial_N\nonumber\\
 \label{eq_square}
&=2\beta_0(\theta\partial_\theta+\tb\partial_\tb)+2(1-\beta_0\tb\ts)\partial_\ts
\partial_\tb
 \end{align}
and where $\square_{\underline{\theta_r}}$ is defined accordingly with the renormalized metric \eqn{eq_renorm_metric}.

To make link with the divergent constants of the previous section, we need to relate the matrix \eqn{eq_param_SUN_2} to the superfield $\Lambda_k$.
We write
\beq
\label{eq_paramtilde}
 \tilde {\cal V}_{k}=\sqrt{Z}\exp\{ig_0\tilde\Lambda_k\}
\eeq
and expand the exponentials in \eqn{eq_param} and \eqn{eq_paramtilde}. Comparing 
with the linear parametrizations, Eqs. \eqn{eq_param_SUN} and 
\eqn{eq_param_SUN_2}, we find $\tilde\Lambda_k=\Lambda_k/\sqrt{Z}+{\cal 
O}(\Lambda_k^2)$. Inserting \eqn{eq_renfields}, \eqn{eq_renconst}, and 
\eqn{eq_paramtilde} in the expressions \eqn{eq_L1} and \eqn{eq_L2} and demanding 
that the effective action written in terms of renormalized quantities be 
finite, we obtain the following relations for the divergent parts of the 
various renormalization factors:
\beq
\label{eq_relations}
\begin{split}
 Z_1&=1/Z_A\,,\\
 \kappa_2/\kappa_1&=1/(Z_\beta Z_A)\,,\\
 \kappa_2\kappa_1&=1/Z_c\,,\\
 \kappa_3&=\xi_0/(Z_\beta Z_\xi Z_c)\,,\\
 \tilde g/\kappa_1&=g_0/(Z_g\sqrt{Z_A})\,,\\
 Z&=Z_\Lambda Z_g^2/Z_\beta\,,
\end{split}
\eeq
as well as the constraint
\beq
\label{eq_constr}
 Z_h=Z_\beta Z_c\,,
\eeq
where we used \Eqn{eq_symreprel}.

To end this section, we mention that the above results generalize those of Ref. \cite{Serreau:2012cg} corresponding to the case of the Landau gauge, $\xi_0=0$. As shown there and rederived in detail in Appendix \ref{appsec_Landau}, in that case the number of independent renormalization factors is reduced from 6 to 3 thanks to further nonrenormalization theorems. In particular, one has $Z_AZ_cZ_\beta=Z_g\sqrt{Z_A}Z_c=Z_\Lambda/Z_c=1$.

\section{Feynman rules}
\label{sec_frules}

The superfield formalism makes transparent the consequences of the 
supersymmetries---the isometries of the curved Grassmann space---for loop 
diagrams. Here, we employ the exponential parametrization \eqn{eq_param}. 
Expanding the action \eqn{eq_action2} in powers of the (super)fields 
$\Lambda_k$, we obtain the vertices of the theory. We work in 
Euclidean momentum space. Because of the curvature of the Grassmann subspace, it 
is of no use to introduce Grassmann Fourier variables. Inverting the quadratic 
part of the action to obtain the free two-point correlators therefore requires a 
bit of Grassmann algebra. The various correlators in the $(A,c,\cb,h)$ sector 
read
 \begin{equation}
  \label{eq_propagAA}
   \left[ A^a_\mu(-p)\,A^b_\nu(p)\right]=\delta^{ab}\left\{\frac{P_{\mu\nu}^T(p)}{p^2+n\beta_0}+\frac{\xi_0P_{\mu\nu}^L(p)}{p^2+\beta_0\xi_0}\right\},
\end{equation}
where $P_{\mu\nu}^L(p)=p_\mu p_\nu/p^2$ and $P_{\mu\nu}^T(p)=\delta_{\mu\nu}-p_\mu p_\nu/p^2$,
\beq
   \label{eq_propagcc}
\left[ c^a(-p)\,\cb^b(p)\right]=\frac{\delta^{ab}}{p^2+\beta_0\xi_0},
\eeq
\beq
   \label{eq_propaghh}
 \left[ih^a(-p)ih^b(p)\right]=\frac{-\beta_0\delta^{ab}}{p^2+\beta_0\xi_0}
\eeq
and
\beq
   \label{eq_propagAh}
 \left[ih^a(-p)A_\mu^b(p)\right]=\frac{i\delta^{ab}p_\mu}{p^2+\beta_0\xi_0}.
\eeq
Here, the square brackets represent an average with the action (\ref{eq_action2}), with $n$ finite. The correlators \eqn{eq_propagAA}-\eqn{eq_propagAh} assume similar forms as in the CF model with the exception that the square mass term in the transverse part of the gauge-field correlator gets a factor $n$ from the replicated superfield sector.\footnote{The CF model is recovered for $n=1$.} 

The correlator of the superfields $\Lambda_k$ reads 
\begin{equation}
  \label{eq_propagLL}
  \left[\Lambda^a_k(-p,\underline{\theta})\,\Lambda^b_l(p,\underline{\theta}')\right]=\delta^{ab}\!\left[\frac{\delta_{kl}\delta(\underline\theta,\underline\theta')}{p^2+\beta_0\xi_0}+\frac{\xi_0(1+\delta_{kl})}{p^2(p^2+\beta_0\xi_0)}\!\right]\!\!,
\end{equation}
where $\delta(\underline\theta,\underline\theta')=g^{-1/2}(\underline{\theta})\,(\thetab-\thetab')(\theta-\theta')$
is the covariant Dirac delta function on the curved Grassmann space: 
$\int_{\underline{\theta}}\delta(\underline\theta,\underline\theta')f(\underline
{\theta})=f(\underline{\theta}')$. Notice that, for $\xi_0\neq 0$, there is a 
nontrivial correlation between different replicas. Finally, there are nontrivial 
mixed correlators
\beq
  \label{eq_propaghL}
 \left[ih^a(-p)\Lambda_k^b(p,\underline{\theta})\right]=\frac{\delta^{ab}}{p^2+\beta_0\xi_0}
\eeq
and
\beq
  \label{eq_propagLA}
 \left[\Lambda_k^a(-p,\underline\theta)A_\mu^b(p)\right]=\frac{i\xi_0\delta^{ab}p_\mu}{p^2(p^2+\beta_0\xi_0)}.
\eeq

Besides the obvious replica symmetry between the replicas $k\ge2$, the symmetry with the fields ($c,\cb,h$) of the replica $k=1$ is encoded in the structure of the correlator \eqn{eq_propagLL}. It is made explicit by using \Eqn{eq_susy} for the replica $k$, with $U_k=\exp\{ig_0\lambda_k\}$ and writing
\beq
 \Lambda_k^a(p,\underline\ts)=\lambda_k^a(p)+\tb c_k^a(p)+\cb_k^a(p) \ts + \tb\ts ih_k^a(p)+\ldots
\eeq
where the dots stand for terms nonlinear in the fields. Identifying the 
coefficients of the terms $\ts$, $\tb$, and $\tb\ts$ on both sides of 
\Eqn{eq_propagLL}, we obtain $[c^a_k\cb^b_l]=\delta_{kl}[c^a\cb^b]$ and 
$[ih^a_kih^b_l]=\delta_{kl}[ih^aih^b]$. Doing the same exercise with 
\Eqn{eq_propagLA} one finds that the mixed terms 
$[h^a_kA_\mu^b]=0\neq[h^aA_\mu^b]$. The replica $k=1$ which has been singled out 
to factor out the volume of the gauge group has a nontrivial mixing with the 
gauge field.

The interaction vertices are obtained from terms higher than quadratic in the fields. From \Eqn{eq_action2}, it appears clearly that the vertices of the sector ($A,c,\cb,h$) are identical to those of the CF model or, equivalently to those of the CFDJ action. These include the Yang-Mills vertices with three and four gluons as well as a standard gluon-ghost-antighost vertex whose expression depends on whether one employs the nonsymmetric or the symmetric version of the CF action, Eqs. \eqn{eq_action_CFDJ} and \eqn{eq_av_sym}, respectively. In addition, there is a four-ghost vertex whose expression also depends on the choice of the nonsymmetric or symmetric CF action. Finally, in the case of the nonsymmetric formulation, there is a $hc\cb$ vertex.\footnote{An alternative formulation of the theory consists in integrating out the field $h$ explicitly. This generates a quadratic term $(\partial_\mu A_\mu^a)^2$ and renormalizes the four-ghost and the ghost-gluon vertices.} These vertices are well known and we do 
not recall their expressions here. In the following, we use the nonsymmetric version of the theory. 

The vertices of the replicated nonlinear sigma model sector are obtained by 
expanding the exponential \eqn{eq_param} in powers of $\Lambda_k$. In this part of 
the action, being linear in the gauge field $A$, there are vertices with an 
arbitrary number of $\Lambda_k$ legs and either one or zero gluon leg. Vertices 
with no gluon legs involve two (normal or Grassmann) derivatives whereas those with one 
gluon leg come with one normal derivative $\partial_\mu$. Furthermore, since the 
color structure of these vertices only involve the antisymmetric tensor 
$f^{abc}$, there is no cubic $\Lambda_k^3$ vertex: 
$f^{abc}\partial_\mu\Lambda_k^a\partial_\mu\Lambda_k^b\Lambda_k^c=f^{abc}g^{MN}
\partial_N\Lambda_k^a\partial_M\Lambda_k^b\Lambda_k^c=0$. Finally we emphasize 
that such vertices do not couple different replicas and are local in Grassmann 
variables, i.e., they are proportional to 
$\prod_{i+1}^{n-1}\delta(\underline{\ts_i},\underline{\ts_{i+1}})$ where 
$\underline{\ts_i}$ are the couples of 
Grassmann variables associated to the $n$ $\Lambda_k$ legs.\footnote{We recall that each replica comes with its own set of Grassmann variables. We omit the replica index $k$ on the latter for simplicity.}    
As an example, the lowest order vertex with $\Lambda_k$ legs, coming from the $A_\mu\partial_\mu\Lambda_k\Lambda_k$ term in the action, reads
\bea
 &&\hspace{-.2cm}\frac{\delta_\theta}{\delta\Lambda_k^a(p_1,\underline\theta)}\frac{\delta_{\theta'}}{\delta\Lambda_l^b(p_2,\underline\theta')}\frac{\delta}{\delta A^c_\mu(p_3)}S\nn
\label{eq_vertex}
 &&\hspace{-.2cm}=i\frac {g_0}4f^{abc}\delta_{kl}(2\pi)^d\delta^{(d)}(p_1+p_2+p_3)\delta(\underline\ts,\underline\ts')(p_1-p_2)_\mu\,.\nn
\eea

For one-loop calculations, only the cubic and quartic vertices are needed. 
These include all the vertices of the CF model described above, the cubic 
$A\Lambda_k^2$ vertex \eqn{eq_vertex}, as well as the quartic vertices  
$A\Lambda_k^3$ and $\Lambda_k^4$. The expression of the latter is quite 
cumbersome and we do not give it here. In fact, at one loop, those quartic 
vertices appear in tadpole diagrams for the $\Lambda$-$\Lambda$ 
and the $A$-$\Lambda$ self-energies, as depicted in Figs. 
\ref{fig_lambda_lambda} and \ref{fig_A_lambda}, respectively. 

To end this section, we mention that we recover the Feynman rules of 
\cite{Serreau:2012cg} for $\xi_0=0$. In that case the superfield propagator 
\eqn{eq_propagLL} is local in Grassmann space which leads to dramatic 
simplifications. In particular, closed loops involving the superfields 
$\Lambda_k$ vanish and the latter thus effectively decouple in the calculation 
of gauge-field and/or ghost correlators. It follows that the  $\xi_0=0$ is 
perturbatively equivalent to the corresponding (Landau gauge) CF model. This is not the case 
for $\xi_0\neq0$. The superfields do not decouple and the theory is not 
equivalent to the CF model.

\section{Renormalization at one loop}
\label{sec_renor1loop}

In this section, we illustrate the renormalizability of the theory by computing 
the divergent parts of the various vertex functions at one-loop order using the 
Feynman rules described above. We explicitly check that the six renormalization 
factors introduced in Sec.~\ref{sec_renorm_pert} are enough to make the theory 
finite. We work with renormalized fields and parameters\footnote{Grassmann 
variables are understood as renormalized ones throughout this section, see 
\eqn{eq_Grass_renorm}.} and employ dimensional regularization with 
$d=4-\varepsilon$.

We introduce the following notation for renormalized two-point vertex functions in momentum space:
\beq
 \left.\frac{\delta^2\Gamma}{\delta\varphi_1^a(p)\delta\varphi_2^b(-p)}\right|_0=\delta^{ab}\Gamma^{(2)}_{\varphi_1\varphi_2}(p),
\eeq
where $\varphi_{1,2}$ denote renormalized fields in the sector 
$(A_r,c_r,\cb_r,ih_r)$ and the derivative on the left-hand side is evaluated at 
vanishing sources. Here, we explicitly extract a trivial color factor. We use a 
similar definition and notation for vertex functions involving the superfields 
$\Lambda_{r,k}$, which now involves a covariant functional derivative 
$\delta/\delta\varphi^a(p)\to\delta_\theta/\delta\Lambda_{r,k}^a(p,
\underline\theta)$, as defined in \eqn{eq_def_sr}.

\begin{figure}[tb]
  \centering
  \includegraphics[width=.17\linewidth]{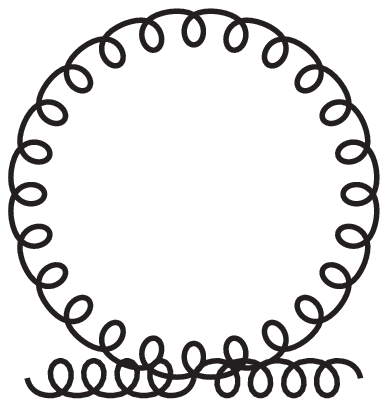}\quad
  \includegraphics[width=.255\linewidth]{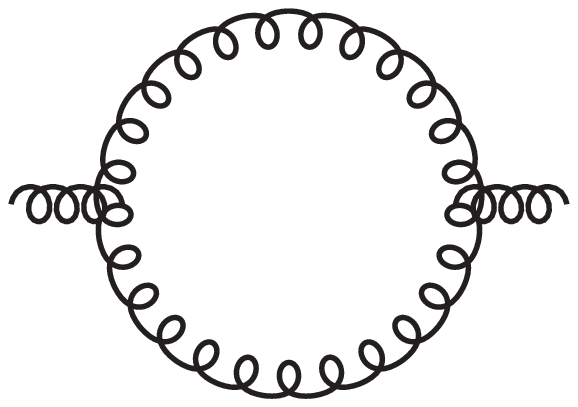}\quad
  \includegraphics[width=.27\linewidth]{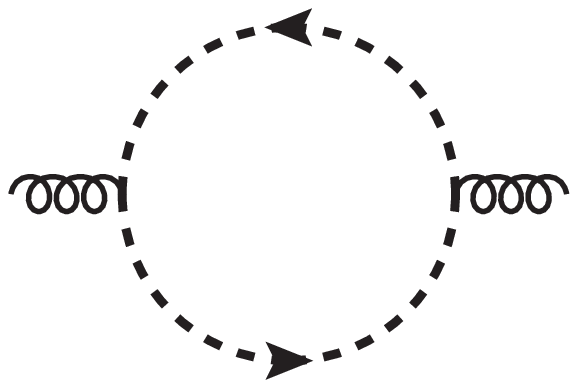}\vspace{.3cm}
  \includegraphics[width=.27\linewidth]{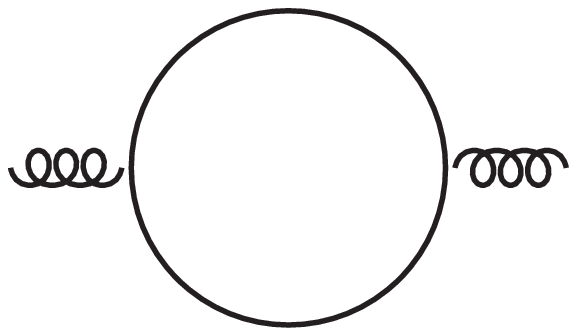}\quad
  \includegraphics[width=.27\linewidth]{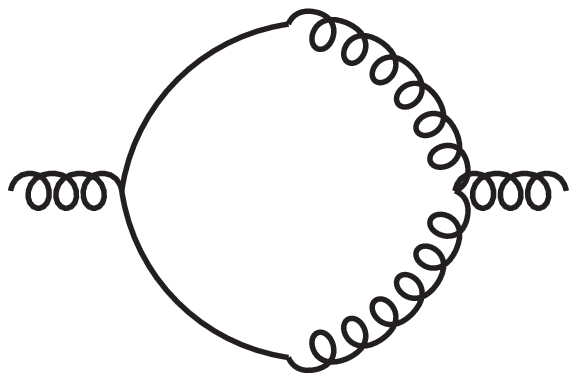}
  \caption{One-loop diagrams for the vertex $\Gamma^{(2)}_{AA}$. We use the standard graphical conventions for the gluon (wiggly) and ghost (dashed) lines. The plain line represents the superfield correlator \eqn{eq_propagLL}. The second diagram on the second line involves a mixed $A$-$\Lambda$ correlator \eqn{eq_propagLA}.}
  \label{fig_AA}
\end{figure}
\begin{figure}[b]
  \centering
  \includegraphics[width=.27\linewidth]{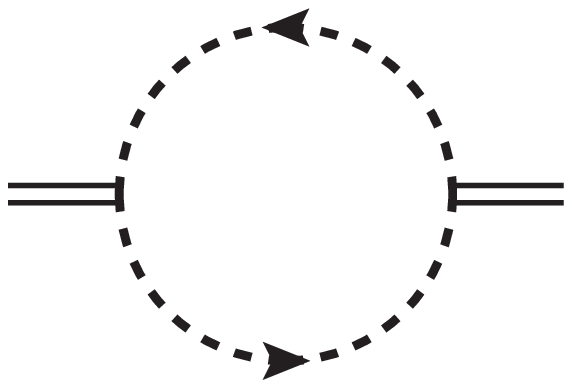}\qquad\qquad
  \includegraphics[width=.27\linewidth]{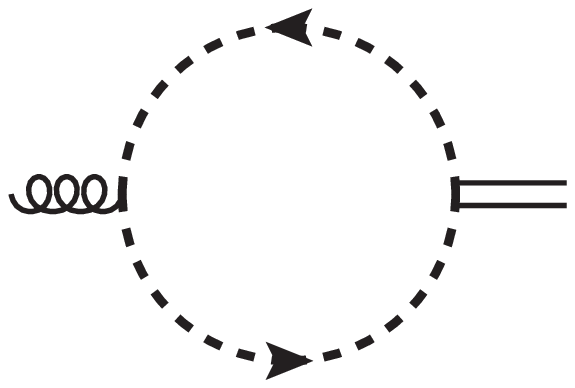}
  \caption{One-loop diagrams for the vertices $\Gamma^{(2)}_{ih\,ih}$ (left) and $\Gamma^{(2)}_{A\,ih}$ (right). Double lines stands for the field $ih$.}
  \label{fig_h_h}
\end{figure}
The one-loop diagrams contributing to the gluon self-energy are depicted in Fig. \ref{fig_AA}. There are the usual diagrams of the CF model plus two diagrams involving the superfield sector. Let us illustrate the calculation of Feynman diagrams with superfield loops on the example of the $\Lambda_k$-loop diagram. Using the expression of the correlator \eqn{eq_propagLL} and the vertex \eqn{eq_vertex}, its contribution to $\delta^{ab}\Gamma_{A_\mu A_\nu}(p)$ reads 
\begin{align}
& \frac{g^2}{8}f^{acd}f^{bef}\!\int\!\frac{d^dq}{(2\pi)^d}(q-r)_\mu(q-r)_\nu\nn
&\quad\times\sum_{k,l=2}^n\int_{\underline\ts,\underline\ts'} \Big[\Lambda_k^c(-q,\underline\ts)\Lambda_l^f(q,\underline\ts')\Big]\Big[\Lambda_k^d(-r,\underline\ts)\Lambda_l^e(r,\underline\ts')\Big]\nn
&=-\delta^{ab}\frac{g^2N}{2}\beta\xi(n-1)\nn
\label{eq_lastline}
&\quad\times\!\int\!\frac{d^dq}{(2\pi)^d}\frac{(q-r)_\mu(q-r)_\nu}{q^2(q^2+\beta\xi)(r^2+\beta\xi)}\left\{1+\frac{n+2}{4}\frac{\beta\xi}{r^2}\right\}\nn
\end{align}
where  $r=p- q$ and where we used $f^{acd}f^{bcd}=N\delta^{ab}$. In computing the Grassmann structure, we used $\delta(\underline\ts,\underline\ts)=0$ and $\int_{\underline\ts}1=\beta$. The UV divergent piece of the resulting momentum integral is obtained as a simple pole in $1/\varepsilon$. \begin{figure}[t]
  \centering
  \includegraphics[width=.17\linewidth]{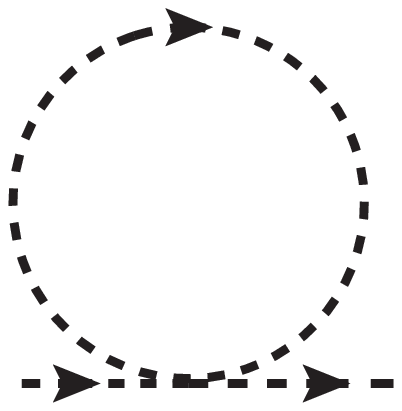}\quad
  \includegraphics[width=.27\linewidth]{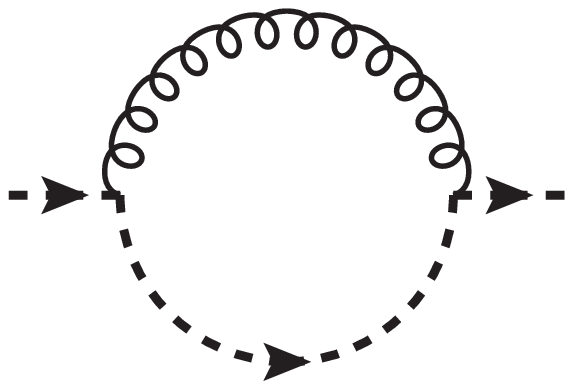}\quad
  \includegraphics[width=.27\linewidth]{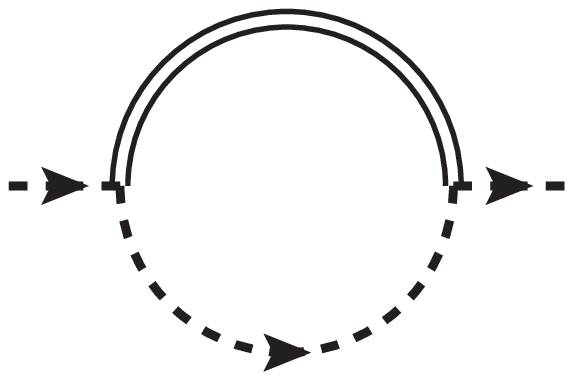}\vspace{.3cm}
  \includegraphics[width=.27\linewidth]{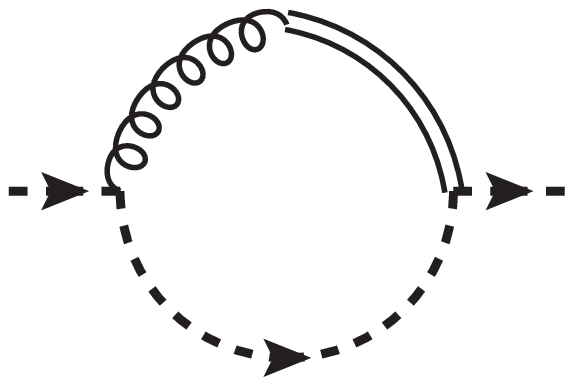}\quad
  \includegraphics[width=.27\linewidth]{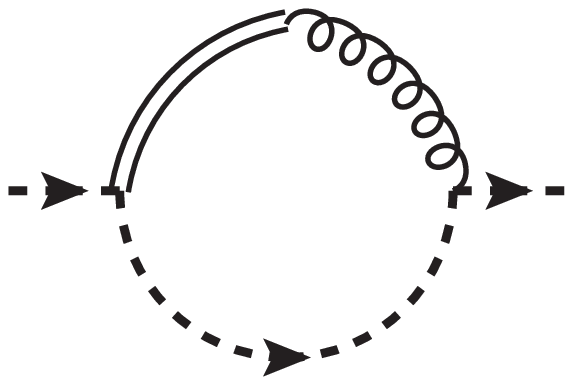} 
  \caption{One-loop diagrams for the vertex $\Gamma^{(2)}_{c\cb}$. The last two diagrams involve the mixed $h$-$A$ correlator \eqn{eq_propagAh}.}
  \label{fig_ccb}
\end{figure}
The second term in brackets on the last line of \eqn{eq_lastline} is UV finite. The divergent contribution of this loop diagram is thus proportional to the number of replicated nonlinear sigma model fields $n-1$. Moreover, the superficial degree of divergence of the integral being zero, its divergent part is a constant and can be obtained by simply setting $p=0$. It reads   
\beq
 -\delta^{ab}\frac{g^2N}{16\pi^2\varepsilon}(n-1)\beta\xi\delta_{\mu\nu}.
\eeq
This is a divergent contribution to the gluon square mass. Each replica $k\ge2$ contributes the same. There is a similar contribution from the ghost loop, hence from the replica $k=1$, which reads 
\beq
 -\delta^{ab}\frac{g^2N}{16\pi^2\varepsilon}\beta\xi\delta_{\mu\nu}.
\eeq
We see that the replica $k=1$ contributes the same as all the other replicas and the resulting contribution is thus proportional to $n\beta$, as expected from the replica symmetry. A simple calculation reveals that other contributions to the gluon mass only come from transverse gluon loops and are thus proportional to the transverse gluon mass $n\beta$. It follows that the total contribution to the gluon mass renormalization is proportional to $n\beta$, as expected from our general proof of renormalizability; see Sec.~\ref{sec_renorm_pert}.

\begin{figure}[t]
  \centering
  \includegraphics[width=.17\linewidth]{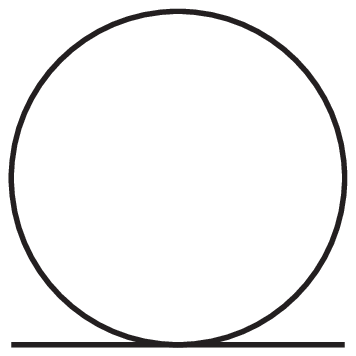}\,\,
  \includegraphics[width=.17\linewidth]{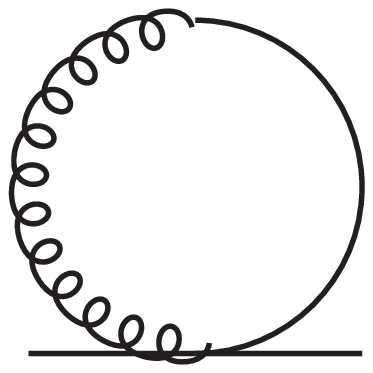}\,
  \includegraphics[width=.27\linewidth]{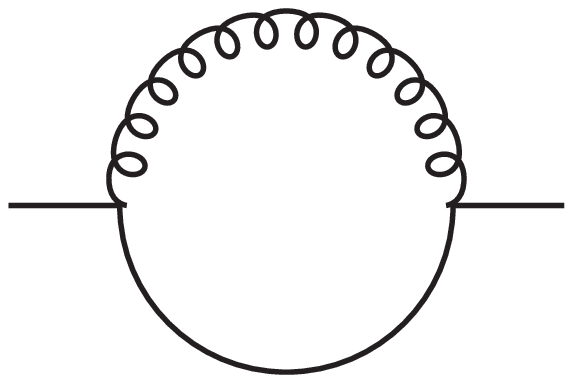}\,
  \includegraphics[width=.27\linewidth]{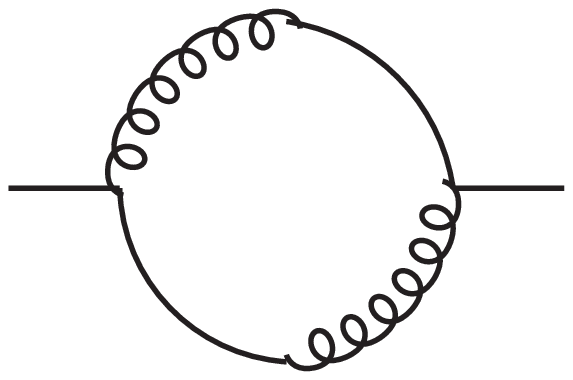}
  \caption{One-loop diagrams for the vertex $\Gamma^{(2)}_{\Lambda\Lambda}$.}
  \label{fig_lambda_lambda}
\end{figure}

\begin{figure}[t]
  \centering
  \includegraphics[width=.17\linewidth]{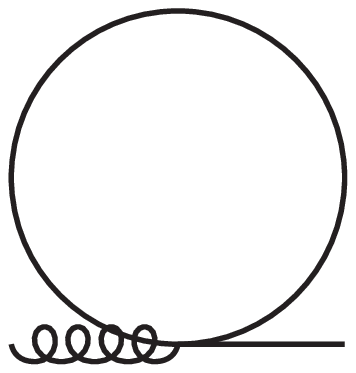}\quad
  \includegraphics[width=.27\linewidth]{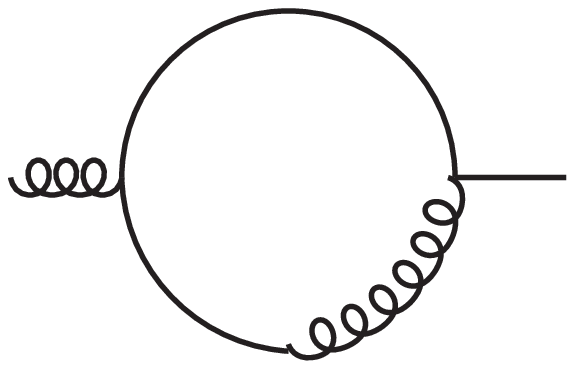}\quad
  \includegraphics[width=.27\linewidth]{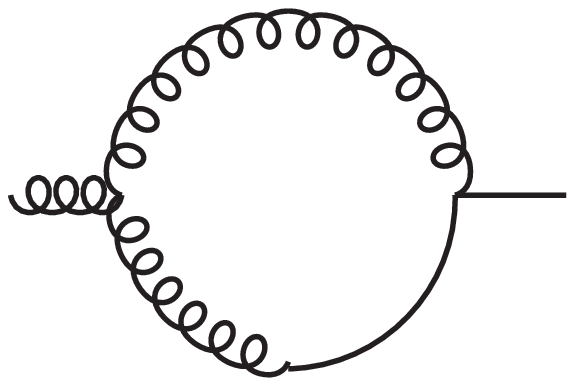}
  \caption{One-loop diagrams for the vertex $\Gamma^{(2)}_{A\Lambda}$.}
  \label{fig_A_lambda}
\end{figure}

The relevant one-loop diagrams are shown in Figs. \ref{fig_AA}-\ref{fig_A_lambda} and can be evaluated along similar lines. Introducing the notation $\kappa=g^2N/8\pi^2\varepsilon$, we obtain, for the divergent parts of the two-point vertex functions,
 \beq
 \label{eq_Gamma_AA}
 \begin{split}
 \Gamma_{A_\mu A_\nu}^{(2){\rm div}}(p)&=n\beta\delta_{\mu\nu}Z_AZ_\beta\left\{1+\kappa\frac{3+\xi}{4}\right\}\\
 &+p^2P_{\mu\nu}^T(p)Z_A\left\{1-\kappa\left(\frac {13}6-\frac {\xi}2\right)\right\},
 \end{split}
\eeq
\beq
 \Gamma_{ih\, ih}^{(2){\rm div}}(p)=-\xi 
Z_hZ_\xi\left(1+\kappa\frac{\xi}{4}\right),
\eeq
\beq
  \Gamma_{ihA_\mu}^{(2){\rm div}}(p)=ip_\mu\sqrt{Z_AZ_h}\left(1+\kappa\frac{\xi}{4}\right),
 \eeq
\beq
 \label{eq_Gamma_cc}
  \Gamma_{c\cb}^{(2){\rm div}}(p)=p^2Z_c\left(1-\kappa\frac{3-\xi}{4}\right)+\beta\xi Z_c Z_\beta Z_\xi\left(1+\kappa\frac{\xi}{4}\right),
\eeq
\beq
  \Gamma_{\Lambda_kA_\mu}^{(2){\rm div}}(p,\underline\theta)=-ip_\mu\sqrt{Z_AZ_\Lambda Z_\beta}\left(1+\kappa\frac{\xi}{6}\right),
 \eeq
 and
\beq
\label{eq_Gamma_LL}
 \begin{split}
  \hspace{-.2cm}\Gamma_{\Lambda_k\Lambda_l}^{(2){\rm div}}(p,\underline\theta,\underline\theta')&=p^2\delta_{kl}\delta(\underline\theta,\underline\theta')Z_\Lambda\left\{1-\kappa\left(\frac34-\frac{\xi}{12}\right)\right\}\\
  &+\frac{\xi}2\delta_{kl}\square_{\underline\theta}\delta(\underline\theta,\underline\theta')Z_\Lambda Z_\xi Z_\beta\left\{1+\kappa\frac{\xi}{12}\right\}.
\end{split}
 \eeq
Here, we used the definition \eqn{eq_square} for the Laplace operator on the curved Grassmann space as well as the identity
$ \square_{\underline\theta}\delta(\underline\theta,\underline\theta')=-2+2\beta\delta(\underline\theta,\underline\theta').$
The divergent parts of the renormalization factors are easily obtained as
 \bea
 \label{eq_ct1}
  Z_A&=&1+\kappa\left(\frac{13}{6}-\frac{\xi}{2}\right),\\
  Z_c&=&1+\kappa\left(\frac{3}{4}-\frac{\xi}{4}\right),\\
  Z_\beta&=&1-\kappa\left(\frac{35}{12}-\frac{\xi}{4}\right),\\
  Z_\xi&=&1+\kappa\left(\frac{13}{6}-\frac{\xi}{4}\right),\\
 \label{eq_ct5}
  Z_\Lambda&=&1+\kappa\left(\frac{3}{4}-\frac{\xi}{12}\right).
\eea
We verify \Eqn{eq_constr} at this order of approximation:
\beq
 Z_h=Z_\beta Z_c=1-\kappa\frac{13}{6}.
\eeq
The nine divergent structures of Eqs. \eqn{eq_Gamma_AA}-\eqn{eq_Gamma_LL} are renormalized by the five independent counterterms \eqn{eq_ct1}-\eqn{eq_ct5}. Observe, in particular, that nontrivial correlations between different replicas do develop but are UV finite. The contrary would spoil the renormalizability of the theory. The remaining renormalization constant $Z_g$ can be determined from the ghost-gluon vertex $Ac\cb$. At one-loop order, there is only one diagram involving superfields contributing to the latter, which is trivially finite. Therefore, the calculation of the divergent contribution follows the corresponding one in the CF model and the renormalization factor can be taken from the existing literature, see, e.g., Refs. \cite{Gracey:2001ma,Kondo:2001tm}:\footnote{The only difference is the mass term $n\beta$ for the transverse part of the gluon correlator. This does not affect the divergent contribution.}
\beq
\label{eq:Zg}
   Z_g=1-\kappa\frac{11}{6}.
\eeq

We notice that the divergent parts of the renormalization factors are 
independent of $n$ at one loop. We thus recover the expressions of 
the independent factors $Z_A$, $Z_c$, $Z_\beta$, $Z_\xi$, and $Z_g$ of the CF 
model (which, we recall, corresponds to $n=1$), where $Z_\beta$ is identified to 
the square mass renormalization factor; see 
\cite{Gracey:2001ma,Kondo:2001tm}.\footnote{It follows that the 
nonrenormalization theorems proved in \cite{Wschebor:2007vh,Tissier_08} for the 
CF model are satisfied at one loop: with the present definitions of the 
renormalization factors, they read, for the divergent parts, 
$Z_AZ_cZ_\beta=Z_g\sqrt{Z_A}Z_c=Z_A^2/Z_\xi^2$. Note that in Refs. 
\cite{Gracey:2001ma,Wschebor:2007vh,Tissier_08} the 
renormalization factor $Z_\xi$ is defined as $Z_A/Z_\xi$.} We do not see any 
reason for this trivial $n$ dependence to hold beyond one loop and we expect 
explicit differences with the CF model to arise at higher loop orders. We stress 
that such differences already 
arise at one-loop order in the finite parts of the vertex functions. Similarly, the factor $Z_\Lambda$ is a specific feature of the present theory which, contrarily to the CF model, is an actual gauge-fixed version of Yang-Mills theories. It is interesting to relate it to the normalization $Z$ of the SU($N$) matrix superfields, see \Eqn{eq_oneZ}. The last relation \eqn{eq_relations} gives, at one loop,
\beq
   Z=1-\kappa\frac{\xi}{3}.
\eeq
Again, we check that, in the Landau gauge, we recover the results of \cite{Serreau:2012cg}, as discussed in Appendix \ref{appsec_Landau}: $Z=1$ and $Z_\Lambda=Z_c$.

 { As a final comment, we mention that in a wide variety of renormalization schemes, one can deduce the beta functions of the theory in the UV from the divergent parts of the counterterms, obtained here at one loop. In particular, from \Eqn{eq:Zg}, we recover the universal one-loop beta function for the coupling constant for any finite value of $n$. Therefore, the presence of replicated scalar fields does not affect the asymptotic freedom of the theory, as expected since these fields arise from a particular  gauge-fixing procedure. This happens because the replicated scalar fields come together with replicated ghost and antighost fields which cancel their contribution to the beta function. Just as the renormalizability of the theory described here, this is a consequence of the supersymmetry of the action \eqn{eq_av_superfield_bis}. }

\section{Summary and perspectives}
\label{sec_sum}

We have proposed a formulation of a class of nonlinear covariant gauges as an extremization 
procedure, which has good properties for {  the purpose of numerical minimization techniques. It is of great interest to investigate its possible lattice implementation, e.g., along the lines of Refs.~\cite{Cucchieri:2009kk,Cucchieri:2010ku,Cucchieri:2011pp}.} Ignoring Gribov ambiguities, 
which is probably a valid procedure in the UV regime, this class of gauge is equivalent to the CFDJ gauges.  
Gribov ambiguities, which corresponding to various possible extrema, can be handled in lattice calculations, 
e.g., by selecting a given extremum (minimum) as done in the minimal Landau gauge.

We have applied the method proposed in Ref. \cite{Serreau:2012cg} to deal 
with Gribov ambiguities in an analytical way, which amounts to averaging over the Gribov copies along each gauge orbit with 
a suitable weight. This lifts the degeneracy 
between the different copies and avoids the usual Neuberger zero problem. We have shown 
that our averaging procedure can be formulated as a local 
action that is perturbatively renormalizable in $d=4$. 
This requires a set of six independent renormalization factors. We have provided 
explicit expressions of the latter at one-loop order in perturbation theory. 

The resulting gauge-fixed theory has the form of the CF model augmented by a nontrivial sector of replicated scalar, ghost, and antighost fields, which can be written as supersymmetric nonlinear sigma models coupled to the gauge field; see \Eqn{eq_action2}. This extends the proposal of Ref. \cite{Serreau:2012cg} away from the particular case of the Landau gauge and provides a more generic framework. For instance, unlike in the Landau gauge, the nonlinear sigma model fields do not decouple in the perturbative calculation of ghost and gluon correlators and the present gauge-fixed version of the Yang-Mills theory exhibits explicit differences with the standard CF model. 

The present proposal opens the way to both lattice and continuum studies of Yang-Mills correlators in covariant gauges away from the Landau gauge. In particular, this allows one to study the gauge dependence of such correlators as well as to gain a more generic---and maybe deeper---understanding of their structure as a function of momentum and, possibly, of the role of Gribov copies. Assuming that, for some range of the weighting parameter $\beta_0$, the average over Gribov copies proposed here is essentially equivalent to randomly picking up a single one,\footnote{A possible picture is, e.g., that the minima of \eqn{eq_func}---the analog of the first Gribov region in the Landau gauge---are deeper in average than other extrema in the landscale of \eqn{eq_func}. In that case, we expect the minima to be essentially equiprobable for $\beta_0$ not to large and other extrema to be suppressed for $\beta_0$ not too small.} we expect the action \eqn{eq_action2} to provide a good starting point for an analytic description of the lattice results. 
An important observation is that the weighting parameter $\beta_0$ provides an effective mass to the various degrees of freedom of the theory, see Sec.~\ref{sec_frules}, and thus regulates infrared fluctuations. As a consequence, we expect perturbation theory to be well defined down to the deep infrared, as in the case of the Landau gauge \cite{Tissier_10}. 

A key point in the scenario described in \cite{Serreau:2012cg} is to absorb the trivial $n$ dependence of the transverse gluon correlator, see \Eqn{eq_propagAA}, in a renormalized square mass parameter $m^2$, defined as $n\beta_0=Z_{m^2}m^2$, in such a way that the latter survives the limit $n\to0$. It appears natural to employ a similar renormalization scheme in the present case in order to recover the Landau gauge scenario in the limit $\xi_0\to0$. Similarly, the ghost mass in \eqn{eq_propagcc} should vanish in this limit. This suggests including an $n$ dependence in the renormalization of the parameter $\xi_0$ as well such that $\beta_0\xi_0$ is finite in the limit $n\to0$ (to be taken before the Landau gauge limit). A simple choice is $\xi_0/n=Z_{\xi'} \xi'$, with $\xi'$ a finite $n$-independent renrormalized gauge-fixing parameter. In this scenario, one has $\beta_0\xi_0\to{\rm const.}$ whereas $\xi_0\to0$ in the limit $n\to0$. Thus, the only departures from the Landau gauge which survive this limit are those involving the product $\beta_0\xi_0$. For instance, a definite prediction is that the ghost correlator, not only its dressing function, is finite at vanishing momentum.\footnote{The possibility of a nonperturbative infrared finite ghost correlator in the Feynman gauge has been discussed in Ref. \cite{Aguilar:2007nf} in the context of Schwinger-Dyson equations.} Similarly, we expect from \eqn{eq_propagAA} the longitudinal component of the gluon correlator to develop a mass gap at zero momentum and to be suppressed by loop effects since its tree-level expression is $\propto\xi_0\to0$.

We stress that the average over Gribov copies proposed here could, in principle, be implemented on the lattice, e.g., along the lines of Ref. \cite{Maas:2013vd}, at least in the case where minima dominate. This offers an interesting possibility to measure new nontrivial quantities such as, in the notations of Sec.~\ref{sec_averaging},  $\overline{\langle A\rangle\langle A\rangle}$ which, as opposed to the correlator $ \overline{\langle A A\rangle}$, concerns correlations between different gauge orbits and might thus bring some information concerning the role of Gribov copies in computing gauged-fixed correlators. This deserves further investigation.

Finally, we mention that since the BRST symmetry of the gauge-fixed Yang-Mills action proposed here is not nilpotent, see \Eqn{eq_algebra}, the standard proof of unitarity does not apply, a situation similar to that of the CF model \cite{Curci:1976bt,deBoer:1995dh,Kondo:2012ri}. It is important to study the question of unitarity in the present context and, in particular, the possible role of the replicated (super)fields and/or of the limit $n\to0$.

\section*{ACKNOWLEDGMENTS}

We are grateful to M.~Pel\'aez, U.~Reinosa, and N.~Wschebor for many useful 
discussions. {  We thank Ph. Boucaud, J.-P. Leroy, and O. P\`ene for sharing their expertise on numerical extremization techniques}.

\appendix

{
\section{Lattice formulation and minimization algorithm}
\label{appsec:min}

Let us briefly discuss how the Los Alamos minimization algorithm \cite{gupta87}, routinely used in lattice calculations in the minimal Landau gauge, can be generalized to the functional \eqn{eq_func}; see also \cite{Serreau:2014xua}. We consider the case of the SU($2$) group for simplicity.\footnote{ A possible generalization to SU($3$) consists in applying the SU($2$) minimization step described below to the three SU($2$) subgroups of SU($3$) alternatively.} Introducing the lattice link variable $W_\mu(x)=\exp\left\{-iag_0A_\mu(x)\right\}$ and the rescaled matrix field $M(x)=a^2g_0^2\eta(x)/2$, where $a$ is the lattice spacing, a simple discretization of the extremization functional \eqn{eq_func} reads \cite{gupta87,Boucaud:2011ug,Serreau:2014xua}, up to an irrelevant constant,
\beq
\label{appeq:minfunc}
 {\cal H}_{\rm latt.}[W,M,U]={\rm Re}\,{\rm tr}\sum_x\left\{M^\dagger(x)U(x)-\sum_{\mu=1}^d W^U_\mu(x)\right\}\!,
\eeq
with $W^U_\mu(x)=U(x)W_\mu(x)U^\dagger(x+\hat\mu)$ the gauge transformed link variable, where $\hat\mu$ denotes a lattice link in the direction $\mu$. 
The second term on the right-hand side of \Eqn{appeq:minfunc} is the
usual discretized version of the Landau gauge extremization
functional. Note that this term, being the trace of a SU(2)
matrix, is bounded from below by $-2dN_{\rm latt.}$, where $N_{\rm latt.}$ is the number of lattice sites. It is easy to prove that the first term
is also bounded from below by $-\sum_x\sqrt {2\, {\rm tr}\, [M^\dagger(x)
M(x)]}$. For a finite lattice and for a given matrix field $M(x)$, the
previous functional therefore admits a minimum.\footnote{With the averaging over $M$ proposed here, see \Eqn{appeq:lattweight} below, the typical value of the lower bound is $\sim -2(d+g_0\sqrt{\xi_0})N_{\rm latt.}$.}

The Los Alamos procedure, that we aim at generalizing in this
appendix, relies on minimizing the functional \eqn{appeq:minfunc} by
making a gauge transformation on one site $x$. An essential property
in this respect is that the functional to be extremized is linear in
the gauge transformation matrix $U(x)$ at each lattice site $x$. As is
well known, this is the case of the second term of
(\ref{appeq:minfunc}), and the first term obviously shares this
property. Under the previous restriction, minimizing the functional
(\ref{appeq:minfunc}) reduces to minimizing
 \beq
\label{appeq:minfunction}
 {\cal H}_{\rm latt.}[W,M,U]={\rm const}-{\rm Re}\,{\rm tr}\left\{U(x)B(x)\right\}\!,
\eeq
where the matrix $B(x)$ can be given the following compact expression:
\beq
\label{appeq:matrixb}
B(x)=-M^\dagger(x)+{\sum_{\mu=-d}^d} W_\mu(x)U^\dagger(x+\hat \mu)
\eeq with $W_{-\mu}(x)\hat =W_\mu^\dagger(x-\hat \mu)$ and
$W_{0}(x)\hat =0$. Note that the matrix $B$ is a generic $2\times 2$
matrix with complex entries because $\eta$, and consequently $M$, are
unrestricted in our implementation, see
Sec.~\ref{sec_gaugefixing}. However, since only the real part of the
trace appears in Eq.~(\ref{appeq:minfunction}), the extremization
procedure is only sensitive to the matrix ($\sigma_{a=1,2,3}$ denotes the Pauli matrices)
\begin{equation}
  \label{eq_hfield}
  C(x)=\frac \openone 4\tr(B(x)+B^\dagger(x))+\frac {\sigma_a}4\tr[\sigma_a(B(x)-B^\dagger(x))],
\end{equation}
which is proportional to a SU(2) matrix. As is well known \cite{deforcrand89}, the minimum of
(\ref{appeq:minfunction}) is attained for $U(x)=U_{\rm min}(x)$ with:
\begin{equation}
  \label{eq_min}
  U_{\rm min}(x)=\frac{C^\dagger(x)}{\sqrt{{\rm det}\, C(x)}}
\end{equation}

The generalization of the standard Los Alamos algorithm for computing
the average of some operator $\mathcal O$ goes as  follows: 
\begin{itemize}
\item For each configuration of the link variables ${W_\mu(x)}$ and of
  the noise field ${M(x)}$, apply the gauge transformation
  $U(x)=U_{\rm min}(x)$ at all even lattice sites (which can be
  treated independently) and $U(x)=1\!\!1$ at all odd lattice
  sites. This decreases the functional \eqn{appeq:minfunc}. Then
  perform another minimization step on odd lattice sites. Repeat these operations until a local minimum is reached.\footnote{ Note that there are equivalent classes of noise fields $M$ which lead to the same extremization problem,  \Eqn{appeq:minfunction}. Indeed, it is always possible to add to $M$ a traceless, real matrix, such that $C$ is not changed in \Eqn{eq_hfield}. We stress, however, that we extremize the functional \eqn{appeq:minfunc} with respect to $U$ at fixed $M$ such that the existence of such equivalent classes has no influence on the minimization procedure. Our formulation relies on taking into account all the matrices belonging to the equivalence class and averaging over them with possibly different weights.}
\item Compute the operator $\mathcal O$ for this gauge configuration,
  and repeat the previous point with another matrix field $M$. Perform
  the average over the matrix field $M$ with weight, see \Eqn{eq_p_eta},
\beq \label{appeq:lattweight}
\hspace{1cm}{\cal P}_{\rm latt.}[M]=\exp\left\{-\frac{1}{\xi_0g_0^2}{\rm
      tr}\sum_x\left[M^\dagger(x) M(x)\right]\right\}.  
\eeq
\item Finally, average over gauge links with the (discretized) Yang-Mills action.
\end{itemize}
 
Of course, the above considerations do not guarantee that the
algorithm converges fast enough for actual
implementations,\footnote{In practice, it may be important to couple
  the previous procedure with some Fourier acceleration methods; see \cite{davies88}.} but our point here is to demonstrate on this particular example that standard algorithms used in the Landau gauge may be easily generalized to the functional \eqn{appeq:minfunc}. The Los Alamos minimization step is the basis of more refined methods such as the (stochastic) over-relaxation algorithm \cite{Cucchieri:1995pn}. We hope this will motivate lattice studies of the present proposal. 
}

\section{Replica symmetry}
\label{appsec:replica}

There is an obvious permutation symmetry among the replicas $k>2$. The latter 
guarantees for instance that $Z_2$ and $ Z_3$ do not depend on $k$; see 
\Eqn{eq_symreprel}. There is also a less obvious permutation symmetry between 
the replicas $k>2$ and $k=1$, which has been (arbitrarily) singled out to factor 
out the volume of the gauge group. To exploit this symmetry we employ a 
parametrization of $\tilde{\cal V}_{k}$ similar to \eqn{eq_susy}
\beq
 \tilde {\cal V}_{k}=\sqrt{Z}\exp\left\{i\tilde g\!\left( \bar 
C_{k}\theta+\bar\theta  C_{k}+\bar\theta\theta\hat H_{k}\right)\right\} \tilde 
U_{k}.
\eeq
with $\hat H_{k}^a=i  H_{k}^a+\frac {\tilde g}2f^{abc}  \bar C^b_{k}  C_{k}^c$ 
and $\tilde U_{k}^\dagger \tilde U_{k}=\openone$. Here, we introduced the fields 
($C_{k},\bar C_{k},H_{k},\tilde U_{k}$) in order to take into account a possible 
renormalization between the bare fields introduced in \eqn{eq_susy} and the 
variables of the effective action $\Gamma$. A simple calculation leads to
\begin{align}
 \int_{\underline\theta}{\cal L}_2&=\frac{Z_2Z^2}{\kappa_1}\bigg\{\frac{\beta_0}{2\kappa_1}
(A_\mu^a)^2-iA_\mu^a\partial_\mu  H_{k}^a \nn
&+\partial_\mu \bar C_{k}^a\left(\kappa_1\partial_\mu   C_{k}^a+\tilde gf^{abc} 
A_\mu^b  C_{k}^c\right)\bigg\}_{A=A^{\tilde U_{k}}}\nn
&+Z_3Z^2\bigg\{\beta_0 \bar C_{k}^a  C_{k}^a+\frac{(  H_{k}^a)^2}{2}\nn
&-\frac 
{\tilde g}2 f^{abc}i  H_{k}^a \bar C_{k}^b  C_{k}^c-\frac{\tilde g^2}4(f^{abc} 
\bar C_{k}^b  C_{k}^c)^2\bigg\}.
\end{align}
To make contact with the original fields ($c_{k},\cb_{k},h_{k}$)  to which the 
replica symmetry applies, we introduce possible renormalization factors as
\bea
   C^a_{k}&=& \hat Z_c c^a_{k}+\ldots\nn
   \label{eq_ZZZ}
   {\bar C}^a_{k}&=& \hat Z_c \cb^a_{k}+\ldots\\
 i  H^a_{k}&=& \hat Z_hih^a_{k}+ \hat Z_A\partial_\mu A_\mu^a+ g_0\hat 
Z_{c\cb}f^{abc}\cb_{k}^bc_{k}^c+\ldots\nonumber
\eea
where the dots stand for terms involving $\lambda_{k}$ and nonlocal 
contributions. Here we included all possible local terms having the correct  
dimension, ghost number, and symmetry properties. The replica symmetry guarantees 
that the factors $\hat Z_{c,h,A,c\cb}$ above do not depend on $k$. Inserting 
\eqn{eq_ZZZ} in the equation above, setting $\tilde U_{k}=\openone$, and identifying terms 
involving $(c_{k},\cb_{k},h_{k})$ with the corresponding ones involving 
$(c,\cb,h)$ in \eqn{eq_L1}, we obtain, after some algebra,
\beq
\hat  Z_c= \hat Z_h=1\quad{\rm and}\quad \hat Z_{c\cb}=\hat Z_A=0,
\eeq
 as well as the two relations 
\beq
 Z_2Z^2=\kappa_1\kappa_2\quad{\rm and}\quad Z_3Z^2=\kappa_3,
\eeq
which reduce the number of independent renormalization constants to six.
Note that $Z_A=0$ guarantees that there is no term $(\partial_\mu A_\mu^a)^2$. 
The relation $Z_2Z^2=\kappa_1\kappa_2$ guarantees that all replicas contribute 
the same to the gluon mass squared, which thus scales as $n$.

\section{Formulation without superfields}
\label{appsec_AT}

\subsection{Feynman rules}

As mentioned in the main text, we can equivalently formulate the present 
gauge-fixed Yang-Mills action without introducing the superfield formulation.  
This amounts, e.g., to work with the action \eqn{eq_action2} expressed through 
Eqs. \eqn{eq_actionCF_U}-\eqn{eq_av_sym}. Eliminating the fields $\hat h$ and 
$\hat h_k$ through the shifts $\hat h^a\to\hat h^a+\partial_\mu A_\mu^a/\xi_0$, and similarly for $\hat h_k$,
we obtain
\begin{equation}
  S=S_{\rm YM}[A]+S_{\rm gf}[A,c,\cb]+\sum_{k=2}^nS_{\rm 
gf}[A^{U_k}\!,c_k,\cb_k],
\end{equation}
where
\beq
  \begin{split}
 S_{\rm 
gf}[A,c,\cb]=\int_x&\bigg\{\frac{\beta_0}{2}(A_\mu^a)^2+\frac{(\partial_\mu 
A_\mu^a)^2}{2\xi_0}\\
 &+\frac 12\Big(\partial_\mu \cb^aD_\mu c^a+ D_\mu\cb^a \partial_\mu c^a\Big)\\
 &+\beta_0\xi_0\cb^a c^a-\frac{g_0^2\xi_0}{8}(f^{abc}\cb^b c^c)^2\bigg\}.
  \end{split}
\eeq

To obtain the Feynman rules, we parametrize the SU($N$) matrix fields $U_k$ as
\beq
 U_k=\exp\left\{ig_0\lambda_k\right\}
\eeq
and expand in powers of $\lambda_k$. The free two-point correlators are given by 
Eqs. \eqn{eq_propagAA} and \eqn{eq_propagcc} and 
\begin{align}
\left[c_k^a(-p)\cb_l^b(p)\right]&=\delta^{ab}\frac{\delta_{kl}}{p^2+\beta_0\xi_0
},\\
\left[\lambda_k^a(-p)\lambda_l^b(p)\right]&=\delta^{ab}\frac{\xi_0(1+\delta_{kl}
)}{p^2(p^2+\beta_0\xi_0)},\\
\left[\lambda_k^a(-p)A_\mu^b(p)\right]&=\delta^{ab}\frac{i\xi_0p_\mu}{
p^2(p^2+\beta_0\xi_0)}.
\end{align}
The vertices are obtained in a straightforward manner.

\subsection{One-loop results}

We have computed explicitly the divergent parts of the various two-point vertex 
functions at one-loop order. The number of vertices---and thus of diagrams---is 
considerably larger than with the superfield formulation but the calculation is 
straightforward. We introduce the renormalized fields
\beq
c_k=\sqrt{Z_c}c_{r,k}\,,\quad\cb_k=\sqrt{Z_c}\cb_{r,k}\,,\quad\lambda_k=\sqrt{
Z_\lambda}\lambda_{r,k},
\eeq
where the replica symmetry implies that the factor $Z_c$ is the one already 
introduced in \eqn{eq_renfields} and that $Z_\lambda$ does not depend on $k$. 
We obtain, for the divergent part of the gluon two-point vertex,
 \beq
 \label{appeq_Gamma_AA}
 \begin{split}
 \Gamma_{A_\mu A_\nu}^{(2){\rm 
div}}(p)&=n\beta\delta_{\mu\nu}Z_AZ_\beta\left\{1+\kappa\frac{3+\xi}{4}\right\}
\\
 &+p^2P_{\mu\nu}^T(p)Z_A\left\{1-\kappa\left(\frac {13}6-\frac 
{\xi}2\right)\right\}\\
&-\frac{n}{\xi}p^2P_{\mu\nu}^L(p)\frac{Z_A}{Z_\xi}\left\{1+\kappa\frac{\xi}
4\right\}.
 \end{split}
\eeq
The first two lines are identical to \Eqn{eq_Gamma_AA} and the last line is the 
renormalization of the $(\partial_\mu A_\mu^a)^2/\xi$ term in the formalism with 
the $h$ fields integrated out. One readily checks that the renormalization 
factors obtained in Sec.~\ref{sec_renor1loop} cancel the divergences in 
\eqn{appeq_Gamma_AA}.

The ghost vertex function is unchanged as compared to the previous calculation 
in Sec.~\ref{sec_renor1loop}. We thus reproduce \Eqn{eq_Gamma_cc}:
\beq
  \Gamma_{c\cb}^{(2){\rm 
div}}(p)=p^2Z_c\left(1-\kappa\frac{3-\xi}{4}\right)+\beta\xi Z_c Z_\beta 
Z_\xi\left(1+\kappa\frac{\xi}{4}\right)\!,
\eeq
which is finite. The calculation of the replicated ghost vertex function 
involves the same diagrams as those of $\Gamma_{c\cb}^{(2)}$ plus some loops 
involving the fields $\lambda_k$. It is a nontrivial check that the latter 
exactly cancel out (not only their divergent parts) as expected from the replica 
symmetry:
\beq
 \Gamma^{(2)}_{c_k \cb_l}(p)=\delta_{kl}\Gamma_{c\cb}^{(2)}(p).
\eeq

The two-point vertices involving the fields $\lambda_k$ read
\beq
 \label{appeq_Gamma_lA}
 \begin{split}
  \Gamma_{\lambda_kA_\mu}^{(2){\rm div}}(p)&=-ip_\mu\beta 
Z_\beta\sqrt{Z_AZ_\lambda} \left(1+\kappa\frac{\xi}{6}\right)\\
  &-ip_\mu\frac{p^2}{\xi}\frac{\sqrt{Z_A 
Z_\lambda}}{Z_\xi}\left\{1-\kappa\left(\frac34-\frac{\xi}6\right)\right\}
\end{split}
\eeq
 and  
\beq
 \label{appeq_Gamma_ll}
 \begin{split}
  \Gamma_{\lambda_k\lambda_l}^{(2){\rm div}}(p)&=\delta_{kl}\beta p^2Z_\lambda 
Z_\beta\left\{1-\kappa\left(\frac34-\frac{\xi}{12}\right)\right\}\\
  &+\delta_{kl}\frac{p^4}{\xi}\frac{Z_\lambda}{Z_\xi} 
\left\{1-\kappa\left(\frac32-\frac{\xi}{12}\right)\right\}.
\end{split}
\eeq
We check that the four divergent structures in \eqn{appeq_Gamma_lA} and 
\eqn{appeq_Gamma_ll} are canceled by the factors $Z_A$, $Z_\beta$, and $Z_\xi$ 
determined previously and 
\beq
 Z_\lambda=1+\kappa\left(\frac{11}{3}-\frac{\xi}{3}\right).
\eeq

Finally, the coupling renormalization factor $Z_g$ is unchanged as compared to 
the calculation of Sec.~\ref{sec_renor1loop}. As expected from the general 
analysis of Sec.~\ref{sec_constraining}, the theory can be made finite by 
adjusting six independent renormalization factors. The factor $Z_\Lambda$ of the 
superfield formulation is replaced by $Z_\lambda$ in the present one. Those two 
factors are not independent as we now show.

\subsection{Relation to superfield formalism}

We wish to relate the formulation of perturbation theory of the present section, 
in terms of the basic fields $\lambda_k$, $c_k$ , $\cb_k$ and $h_k$, to that of 
the main text, in terms of the superfields $\Lambda_k$. Identifying the 
representations \eqn{eq_susy} and \eqn{eq_param} and using the 
Campbell-Haussdorf formula for the product of exponentials, we get
\beq
 \Lambda_k=\lambda_k+\tb c'_k+\cb'_k\ts+\tb\ts \hat h'_k
\eeq
with
\begin{align}
\label{appeq_cprime}
c'_k&=c_k+\frac{ig_0}{2}[c_k,\lambda_k]-\frac{g_0^2}{12}[[c_k,\lambda_k],
\lambda_k]+\ldots\,,\\
\cb'_k&=\cb_k+\frac{ig_0}{2}[\cb_k,\lambda_k]-\frac{g_0^2}{12}[[\cb_k,\lambda_k]
,\lambda_k]+\ldots
\end{align}
and
\beq
\begin{split}
\label{appeq_hprime}
 \hat h'_k&=\hat h_k+\frac{ig_0}{2}[\hat h_k,\lambda_k]\\
 &-\frac{g_0^2}{12}\left\{[[\hat 
h_k,\lambda_k],\lambda_k]-[[\cb_k,c_k],\lambda_k]\right\}+\ldots\,,
\end{split}
\eeq
where the dots denote higher order nonlinear terms. The above relations 
highlight the fact that the superfield $\Lambda_k$ is a nonlinear composite of 
the original fields $\lambda_k$, $c_k$ , $\cb_k$, and $h_k$. They hold for the 
basic fields to be integrated over in the path integral. Instead, the variables 
of the effective action $\Gamma$ correspond to averages of these integration 
variables in the presence of nontrivial sources, e.g., 
$\langle\Lambda_k\rangle$, where the brackets denote the relevant average in 
the presence of sources as in \Eqn{eq_sources}. In the following, we omit the 
brackets for simplicity. At the level of the effective action, we thus need to 
take into account the nontrivial renormalization of the nonlinear composite 
fields \eqn{appeq_cprime}-\eqn{appeq_hprime}. In particular, one has, at linear 
order,
\begin{align}
 c'_k&=\sqrt{Z_c'} c_k+\ldots\,,\\
 \cb'_k&=\sqrt{Z_c'} \cb_k+\ldots\,,\\
 \hat h'_k&=\sqrt{Z_h'} ih_k+\ldots\,,
\end{align}
where the dots stands for nonlinear and/or nonlocal contributions and where the 
composite field renormalization factors $Z_c'$ and $Z'_h$ do not depend on the 
replica index due to the replica symmetry. At linear order, we write
\beq
 \Lambda_k=\lambda_k+\sqrt{ Z_c'}\tb c_k+\sqrt{ Z_c'}\cb_k\ts+\sqrt{ Z_h'}\tb\ts 
ih_k +\ldots
\eeq
or, equivalently, introducing renormalized fields as in Eqs. \eqn{eq_renfields} and 
\eqn{eq_renormL},
\beq
\label{appeq_renL}
\begin{split}
 &\sqrt{\frac{Z_\Lambda}{Z_\beta}}\Lambda_{r,k}=\sqrt{Z_\lambda}\lambda_{r,k}\\
 &+\sqrt{\frac{Z_cZ_c'}{Z_\beta}}\left(\tb_{r} 
c_{r,k}+\cb_{r,k}\ts_r+\sqrt{\frac{Z_h'}{Z_c'}}\tb_r\ts_r ih_{r,k}\right) 
+\ldots\,,
\end{split}
\eeq
where we used $Z_h=Z_\beta Z_c$; see \Eqn{eq_constr}. 

Now it is easy to check that, for the quadratic part of the effective action to 
have the desired expressions in terms of either the renormalized superfields 
$\Lambda_{r,k}$ or the renormalized fields $\lambda_{r,k}$, $c_{r,k}$, 
$\cb_{r,k}$, and $h_{r,k}$, we must have
\beq
  \Lambda_{r,k}=\lambda_{r,k}+\tb_r c_{r,k}+\cb_{r,k}\ts_r+\tb_r\ts_r i 
h_{r,k}+\ldots
\eeq
We conclude that
\beq
 Z_h'=Z_c'
\eeq
and that
\beq
\label{appeq_relation}
 Z_\lambda=\frac{Z_\Lambda}{Z_\beta}=\frac{Z_c Z_c'}{Z_\beta}.
\eeq

The first equality in \Eqn{appeq_relation} is satisfied at 
one loop, see Eqs. \eqn{eq_ct1}-\eqn{eq_ct5}. The second equality can be checked at 
one loop by a direct calculation of the composite field renormalization factor 
$Z'_c$. Using \Eqn{appeq_cprime} and the appropriate Feynman rules, we obtain, 
after a simple calculation,
\beq
 Z_c'=1+\kappa\frac{\xi}{6}.
\eeq
This agrees with the second equality in \eqn{appeq_relation}.

\section{Landau gauge}
\label{appsec_Landau}

The case $\xi_0=0$ studied in \cite{Serreau:2012cg} exhibits various 
simplifications as compared to the general class of gauges studied here. The first 
obvious one is the fact that the $h$ sector does not receive any loop 
corrections, i.e., 
\beq
 \frac{\delta\Gamma}{\delta ih^a}=\frac{\delta S}{\delta ih^a}=\partial_\mu 
A_\mu^a.
\eeq
This can be seen, e.g., by applying a infinitesimal shift $ih\to ih+f$ under the 
defining path integral for $\Gamma$.
In terms of the divergent constants introduced in Sec.~\ref{sec_constraining}, 
this implies that $\kappa_2=1$ or, equivalently~\cite{Dudal:2002pq,Tissier_08},
\beq
\label{appeq_Landau1}
 Z_AZ_\beta Z_c=1,
\eeq
where we used the relation \eqn{eq_constr}.

The next simplification comes from the fact that for $\xi_0=0$, the superfield 
correlator \eqn{eq_propagLL} is ultralocal in Grassmann space, i.e., it is 
proportional to $\delta(\underline\ts,\underline\ts')$ and the mixed correlator 
\eqn{eq_propagLA} vanishes. As pointed out in Ref. \cite{Serreau:2012cg}, since all vertices 
of the theory are also local in Grassmann space (for $\xi_0=0$ there are no term 
involving Grassmann derivatives) it follows that closed loops involving 
superfields are proportional to $\delta(\underline\ts,\underline\ts)=0$. An 
important consequence is that the $\Lambda_k$ sector  effectively 
decouples in the (perturbative) calculation of correlators in the sector 
$(A,c,\cb,h)$ at all orders. The only effect of the superfields $\Lambda_k$ is 
the mass term $n\beta_0$ for the gauge field. 

A first consequence of this dramatic simplification for the renormalizability of 
the theory is that the usual nonrenormalization theorems of the CF model with 
$\xi_0=0$ are valid. One of them is the relation \eqn{appeq_Landau1} above. The 
second one comes from the Taylor theorem, which states that the 
ghost-antighost-gluon vertex in a particular momentum configuration does not 
receive loop corrections \cite{Taylor:1971ff}. It follows that 
\beq
\label{appeq_Landau2}
 Z_g\sqrt {Z_A}Z_c=1.
\eeq
We observe that Eqs. \eqn{appeq_Landau1} and \eqn{appeq_Landau2} together with the last relation in \Eqn{eq_relations} imply that $Z=Z_\Lambda/Z_c$. The second equality in \Eqn{appeq_relation}, thus shows that the normalization $Z$ of the nonlinear sigma model superfields is directly related to the composite field renormalization discussed in Appendix \ref{appsec_AT}: $Z=Z_c'$.

Another consequence of the absence of loops of the superfield is the fact that there can be no loop diagram with only one external $\Lambda_k$ leg. This is easy to show by direct inspection. It follows that those vertices are tree-level exact, i.e., 
\beq
 \left.\frac{\delta_\theta\Gamma}{\delta \Lambda_k^a}\right|_{\Lambda_k=0}=\left.\frac{\delta_\theta S}{\delta \Lambda_k^a}\right|_{\Lambda_k=0} =-\partial_\mu A_\mu^a.
\eeq
Taking into account the rescaling \eqn{eq_Grass_renorm} of Grassmann variables, which implies, together with \eqn{eq_renormL}, that
\begin{align}
 \delta{\cal F}[\Lambda]=\int_{x,\underline\ts}\frac{\delta_\theta{\cal F}}{\delta\Lambda}\delta\Lambda=\sqrt{Z_\beta Z_\Lambda}\int_{x,\underline{\ts_r}}\frac{\delta_{\theta}{\cal F}}{\delta\Lambda}\delta\Lambda_r,
\end{align}
for any given functional ${\cal F}[\Lambda]$, we conclude that $Z_AZ_\Lambda Z_\beta=1$.
When combined with \Eqn{appeq_Landau1}, this gives
\beq
\label{appeq_Landau3}
 Z_\Lambda=Z_c,
\eeq
or, equivalently, $Z=1$. Equations \eqn{eq_ansatz_gamma0}, \eqn{eq_L1}, \eqn{eq_L2}, together with the relations \eqn{eq_symreprel},  \eqn{eq_relations}, \eqn{appeq_Landau1}, \eqn{appeq_Landau2}, and \eqn{appeq_Landau3}, the results of \cite{Serreau:2012cg}. In the Landau gauge, the number of independent renormalization factors is reduced from 6 to 3. The relations \eqn{appeq_Landau1}, \eqn{appeq_Landau2}, and \eqn{appeq_Landau3} are readily checked from the one-loop results of Sec.~\ref{sec_renor1loop} for $\xi=0$.

\end{document}